\documentclass[pra, aps, twocolumn, groupedaddress, showpacs, superscriptaddress]{revtex4-2}
\usepackage{amssymb,amsmath,amsthm,color,graphicx,times,graphicx}
\usepackage{hyperref}
\usepackage{subfigure}
\usepackage{times}
\usepackage{bbold}
\usepackage{color}

\providecommand{\openone}{\leavevmode\hbox{\small1\kern-4.3pt\normalsize1}}

\theoremstyle{plain}

\usepackage{orcidlink}
\theoremstyle{definition}

\usepackage{hyperref}
\hypersetup{
	colorlinks=true,
	linkcolor=blue,
	filecolor=blue,      
	urlcolor=blue,
	citecolor=violet,
	pdfauthor={Abdallah Slaoui, Brahim Amghar and Rachid Ahl Laamara},
	pdftitle={Estimating phase parameters of a three-level system interacting with two classical monochromatic fields in simultaneous and individual metrological strategies},
	pdfcreator={Abdallah Slaoui},
}

\begin{document}
\title{Interferometric phase estimation and quantum resources dynamics in Bell coherent-states superpositions generated via a unitary beam splitter}
\author{Abdallah Slaoui \orcidlink{0000-0002-5284-3240}}\email{Corresponding author: abdallah.slaoui@um5s.net.ma}\affiliation{LPHE-Modeling and Simulation, Faculty of Sciences, Mohammed V University in Rabat, Rabat, Morocco.}\affiliation{Centre of Physics and Mathematics, CPM, Faculty of Sciences, Mohammed V University in Rabat, Rabat, Morocco.}
\author{Brahim Amghar}\affiliation{LPHE-Modeling and Simulation, Faculty of Sciences, Mohammed V University in Rabat, Rabat, Morocco.}\affiliation{Centre of Physics and Mathematics, CPM, Faculty of Sciences, Mohammed V University in Rabat, Rabat, Morocco.}
\author{Rachid Ahl Laamara}\affiliation{LPHE-Modeling and Simulation, Faculty of Sciences, Mohammed V University in Rabat, Rabat, Morocco.}\affiliation{Centre of Physics and Mathematics, CPM, Faculty of Sciences, Mohammed V University in Rabat, Rabat, Morocco.}

\begin{abstract}
Beam splitters are optical elements widely used in modern technological applications to split the initial light beam into a required number of beams and they play a very promising role for generating entangled optical states. Here, a potential scheme is proposed to generate Bell coherent-states superpositions through the action of a beam splitter when a Glauber coherent state is injected on one input mode and vacuum state is incident on the other one. Different quantifiers are used to measure the quantumness in the output state such as concurrence entanglement, entropic quantum discord, quantum coherence, geometric measure of quantum discord, local quantum uncertainty (LQU) and local quantum Fisher information. Thereby, we derive their analytical formulas and focus more on the behavior and bounds of each measure. Besides, we have introduced the notion of "weak measurement-induced LQU" captured by weak measurements as the generalization of normal LQU defined for standard projective measurement, and we investigate the effect of the measurement strength on the estimated phase enhancement if the generated Bell cat states are the probe states in quantum metrology. Our results suggest that the sensitivity of the interferometric phase estimation depends on how strongly one perturbs the probe state and that a weak measurement does not necessarily capture more quantumness in composite system.\par
\vspace{0.25cm}
\end{abstract}
\date{\today}

\maketitle
\section{Introduction}
In today's era, the field of quantum information is developing very fast with the extremely strong development of science and technology. In the information processing, information security have been given top priority \cite{Nielsen2010}. In this emerging field, quantum computing and quantum information transmission are studied by leading theoretical and experimental researchers, as they promise a new revolution in quantum communication techniques \cite{Pirandola2015}. Typically, the Nobel Prize in physics 2022 went to three scientists Aspect, Clauser and Zeilinger for their research related to the quantum field. Their results have paved the way for the application of new technologies as well as for broad areas of research such as quantum computers, quantum networks and quantum communications. The birth of quantum information science occurred since Schrödinger introduced the concept of quantum entanglement (QE) to explain the Einstein-Podolsky-Rosen paradox \cite{Einstein1935}. Later, the ideas of quantum information systems were introduced in 1980 by Manin \cite{Manin1980} and in 1982 by Feynman \cite{Feynman1982}, respectively. Nowadays, such theoretical ideas have so far been initially realized by many economically and technologically powerful countries.\par

Within this context, entangled sources play a very important role in the performance of quantum tasks \cite{Andersen2015} where the class of non-classical states plays a key role in the implementation of quantum protocols such as quantum dense coding \cite{BraunsteinH2000}, quantum key distribution \cite{Ralph1999}, quantum error correction \cite{Braunstein1998} and especially quantum teleportation \cite{Braunstein2000}. Non-classical states having two modes such as the squeezed states \cite{Caves1985} or the coherent states \cite{Agarwal1988} have been proposed. Based on these two non-classical states, many new non-classical states were introduced and their non-classical properties were interesting \cite{Duc2014,Schnabel2017,Clark2016}. In particular, a new family of two-mode non-classical states was introduced based on the technique of adding photons \cite{Hu2013,Hong1999}, reducing photons \cite{Olivares67,Opatrny2000}, adding and deleting photons \cite{Duc2021,Chunqing2000} to both modes of the original state. These states exhibit elevated non-classical properties, high entangled properties and become resources for performing quantum tasks \cite{Hoai2016,Wang2015}. Besides, ever since the introduction of coherent states, many new multi-mode non-classical states have been proposed also by photon addition and deletion techniques that have shown enhanced non-classical properties \cite{Dat2020}. Very recently, using the near-degeneracy of the transverse modes of a linear Paul trap, Jeon and his collaborators \cite{Jeon2023} have proposed an experimentally realized scheme for generating entangled coherent states with two-dimensional motion of a trapped ion. An experimental scheme for generating non-classical multiphoton states through photon subtraction is reported in Ref.\cite{Loaiza2019}. This scheme reveals new mechanisms for controlling the fundamental properties of light, and a family of quantum correlated multiphoton states with tunable average photon numbers and degree of correlation are produced by manipulating the quantum electromagnetic fluctuations of two-mode squeezed vacuum states. An alternative technique is proposed in \cite{Takahashi2010} for the entanglement distillation and generation of a squeezed vacuum state using continuous variable systems. In this scheme, the initial Gaussian entangled state is prepared by splitting the squeezed vacuum by half at the first beam splitter and considering local photon subtraction as non-Gaussian operations. Currently, a variety of devices, including cavity QED \cite{Zheng2000}, beam splitters \cite{Toth2003,Tan1991}, NMR systems \cite{Gershenfeld1997}, have been suggested and experimentally achieved to produce quantum entanglement. Alternatively, quantum entanglement between two modes can be created when using beam splitter, one of the linear optical devices. Its application as an entangler has been the subject of numerous studies for inducing entanglement. Mathematically, the action of beam splitters is described by a unitary transformation linking the input and output fields as a lossless four-port device. The entanglement characteristics of a beam splitter with various input states, including coherent states, Fock states, pure and mixed Gaussian states and squeezed states, have been investigated by Kim and his team \cite{Kim2002}. They hypothesized that at least one of the input fields must be nonclassical in order to create entangled output states of a beam splitter. This has been later demonstrated in \cite{Xiang-Bin2002}. In this work, we investigate in depth the quantum criteria generated via beam splitter using coherent-state input mode. In particular, we consider a Glauber coherent state and a vacuum state at two input modes, respectively.\par
As an extensive resource, quantum correlation directly affects the efficiency and reliability of quantum information processing. For a long time, researchers have always believed that QE is the only type of quantum correlation, so quantum correlation has remained only once in the QE study \cite{MohamedRahman2022,Rahman2022}. However, experiments and theory have confirmed that QE does not include all quantum correlations. In order to describe quantum correlations more completely, Ollivier and Zurek proposed quantum discord (QD) \cite{Ollivier2002} and found that QD is more universal, advantageous and captures a more general quantum correlation than QE. Further studies have shown that QD can not only improve computational speed \cite{Datta2008,Brodutch2012}, but can also be applied to quantum information processing tasks such as remote state preparation \cite{Daki2012,Giorgi2013} and quantum cryptography \cite{Grimaudo2019,Pirandola2014}. QD is not only a kind of information resource, but also a kind of physical resource because the quantum correlation change must be accompanied by the entropy generation of the system and the environment. And more interestingly, some studies have shown that the dynamics of the QD can maintain a constant value for a long time even if the QE suddenly disappears \cite{Wang2012}. A great deal of attention has been devoted to reveal the degree of correlation in quantum systems using different discord-like measures such as Geometric quantum discord (GQD) \cite{Paula2013}, measurement-induced nonlocality \cite{Luo2011}, local quantum uncertainty (LQU) \cite{Girolami2013,Slaoui2018,Slaoui22019}, one-way quantum deficit \cite{Horodecki2005}, quantum dissonance \cite{Modi2010} and local quantum Fisher information (LQFI) \cite{Kim2018,MohamedKhedr2022}.\par
Previous research on QD is based only on the measurement of relatively strong interaction between the system under test and the measuring device, such as the projection measurement, but when the interaction between the system under test and the measuring device is weak, it is required to examine the super-QD \cite{Singh2014,Jing2017}. It has been proven that for a two-qubit quantum system, a weak measurement on one of the subsystems will lead to super-QD. According to the traditional interpretation of quantum mechanics, the measurement of a quantum system causes it to collapse into a new state different from the one before the measurement, i.e. the quantum measurement process will change the state of the measured system in most cases. Nevertheless, weak measurements have been introduced as a theoretical framework \cite{Oreshkov2005} that allows us to explore quantum systems with minimal impact on the investigated system. Quantum correlations under weak measurements are an emerging research topic whose properties differ from those of projective measurements. Recently, some researchers have experimentally simulated the effect of a weak POVM on a nuclear magnetic resonance quantum information processor using weak measurements \cite{Gautam2020}. Here, we found an analytical formula for the LQU obtained via weak measurements for arbitrary qubit-qudit quantum systems, and studied their dynamics in the generated Bell cat states in the weak and strong measurement regions.\par

In several applications, including quantum sensing \cite{Pirandola2018}, quantum imaging \cite{Taylor2013}, and gravitational wave detection \cite{Adhikari2014}, the estimation of the quantum optical phase is a crucial step. The majority of studies are focused on constant phase estimation $\varphi$, for which a Mach-Zehnder interferometer is the most often employed instrument. When using classic resources, the shot noise restricts the estimation's accuracy. This upper bound is frequently referred to as the standard quantum limit, $\varDelta\varphi\propto1/\sqrt{N}$, where $\varDelta^{2}$ is the variance of the measurement outcomes related to each probe and $N$ is the average number of photons in the probe state \cite{Caves1981,Giovannetti2011}. Even greater precision can frequently be obtained with a customary $\sqrt{N}$ improvement by combining the same physical resources with quantum effects like QE or squeezing \cite{Toth2014}. The quantum Fisher information plays a central role in quantum metrology and its inverse provides a lower bound on the statistical estimation error of an unknown parameter embedded in a unitary dynamics \cite{Huelga1997,Hauke2016,McCormick2019}. Therefore, ways to increase the QFI become an intriguing issue in quantum estimation theory. Indeed, the QFI is upper bounded by the variance of a generator as ${\cal F}_{Q}\leqslant4\varDelta^{2}{\cal H}$ with equality for a pure entangled state \cite{Helstrom1969}. Moreover, discord-type correlations beyond QE are necessary and sufficient for quantum enhanced metrology. Hence, the QD measures based on QFI allows us to better understand how QD play a critical role in defining metrological accuracy. Below, we employ the metrological measures of non-classical correlations such as LQU and LQFI to fully establish the role of quantumness in interferometric phase estimation for generated Bell coherent-states superpositions.\par
Our paper is organized as follows. In Sec.\ref{SecI} we derive a general formalism for the arbitrary Glauber coherent state into the beam-splitter input port and explore the way Bell cat states are generated at the beam-splitter output port. In Sec.\ref{SecIII} we briefly present the concepts and definitions of the used quantum criteria such as QD, GQD, QC and LQFI, and introduce a weak measurement-induced LQU as a generalization of normal LQU. The non-classical correlations produced by the effect of a beam splitter on the Glauber coherent state is explained in Sec.\ref{SecIV} using QE, QD and GQD. We also provide the explicit expression of quantum coherence (QC) for Bell cat states under amplitude damping, by employing the concept of the quantum Jensen-Shannon divergence. Here, the interplay of the signal with a vacuum mode in a beam splitter provides for an effective modeling of amplitude damping. Furthermore, if the generated Bell cat states are the probe states, the role of these quantities and the effect of the measurement strength on the estimated phase enhancement are examined. A summary of our results with future directions is done in Sec.\ref{V}.
\section{The Quantum Model: photon loss mechanism of Bell cat states}\label{SecI}
\subsection{Beam splitting transformation}
The beam splitter provides a simple technique to probe the quantum nature of electromagnetic fields through uncomplicated experiments. The study of entangled states has rekindled interest in this device. The way quantum states behave when passing via a beam splitter has been studied by numerous authors \cite{Sanders1992,Sanders1995}. Additionally, the quantum beam splitter network has been used to create multi-particle entangled states in continuous variables \cite{Loock2000} as well as multi-particle entangled coherent states \cite{Wang2002}. The amplitude damping associated to the absorption of photons conveyed in a noisy channel is likewise easily realized in this way.\par
Indeed, the beam splitter is an optical device with two input ports and two output ports that control the interplay of two harmonic oscillators. Here, we first discuss how a beam splitter assigns an input state consisting of a state of interest to be investigated $\left|\psi\right\rangle$ in one input port, and a vacuum state $\left| 0\right\rangle $ in the other port. We consider that the vertical input beam is processed in the vacuum state and that the horizontal input beam contains the state of interest. The transformation law of the quantum states under the effect of the beam splitter, in the Schrödinger picture, is described by the unitary operator ${\cal \hat{B}}\left(\theta\right)$ defined as $\left|\rm out \right\rangle={\cal \hat{B}}\left(\theta\right)\left|\rm int \right\rangle$, with the input state $\left|\rm int \right\rangle=\left|\psi \right\rangle\otimes\left|0\right\rangle$ and the beam splitter operator ${\cal \hat{B}}\left(\theta\right)$ of angle $\theta$ (an element of the $su(2)$ group) is
\begin{equation}
	{\cal \hat{B}}\left(\theta\right) = \exp\left[\frac{\theta}{2}  \left(a_{1} a^{\dagger}_{2} +a^{\dagger}_{1} a_{2}\right)\right],
\end{equation}
where $a_{1}$ and $a_{2}$ (respectively $a^{\dagger}_{1}$ and $a^{\dagger}_{2}$) are the boson-annihilation (respectively creation) operators describing the two input fields. If input state includes a Fock state in the horizontal input beam, i.e. $\left|\psi\right\rangle\equiv\left|n \right\rangle$, and a vacuum Fock state in the other port, the output state is obtained by this transformation
\begin{equation*}
	\left|n \right\rangle\left|0 \right\rangle\xrightarrow{{\cal \hat{B}}\left(\theta\right)} \sum_{p=0}^{n}\left(\begin{array}{c}n\\p
	\end{array}
	\right)^{\frac{1}{2}} T^{p}R^{n-p}\left|p \right\rangle \left|n-p \right\rangle,
\end{equation*}
where the quantities $T$ and $R$ are the transmission and reflection coefficients of the beam splitter, respectively, obeying the normalization condition $|T|^{2}+|R|^{2}=1$, and which are defined in terms of the angle $\theta$ by
\begin{equation}
	T = \cos\left(\theta/2\right), \hspace{2cm}  R = \sin\left(\theta/2\right).
\end{equation}
Then, the state emitted by the beam splitter on an input state $\left| n , 0 \right\rangle$ is given by
\begin{align}
	{\cal \hat{B}}\left(\theta\right) \left| n , 0 \right\rangle= \left(1 +  |\xi|^2
	\right)^{-\frac{n}{2}} \sum_{p=0}^{n} \xi^{p}
	\frac{\sqrt{n!}}{\sqrt{(n - p)!p!}}  \left| p, n - p \right\rangle,
	\label{suncs}
\end{align}
where  the new variable $\xi=R/T$ is defined as the ratio of the transmissivity and the reflectivity of the considered beam splitter. Basically, the input state should be non-classical for a beam splitter to produce entanglement in the output state. If a Glauber coherent state defined as $\left|\alpha\right\rangle= \exp\left[-\frac{|\alpha|^2}{2}\right] \sum_{n=0}^{\infty}
\frac{\alpha^n}{\sqrt{n!}}\left|n\right\rangle$ is injected into one input port of the beam splitter and a vacuum state into the other, then the output state is factorizable with zero entanglement
\begin{align}
	{\cal\hat{B}}\left(\theta\right)\left|\alpha,0 \right\rangle &=\exp\left[-\frac{|\alpha|^2}{2}\right]\sum_{p=0}^{\infty}\frac{\left(\alpha T\right)^{n}}{\sqrt{n!}}\left|n\right\rangle\otimes\notag\\&\sum_{p=0}^{\infty}\frac{\left(\alpha R\right)^{\left(n-p \right)}}{\sqrt{\left(n-p\right)!}}\left|n-p\right\rangle\notag\\&=\left|\alpha T\right\rangle\otimes\left|\alpha R\right\rangle.
\end{align}
If we examine the superposition of the form
\begin{equation}
	\left|\Psi_{\alpha}\right\rangle = \frac{1}{\sqrt{N_{\alpha}}}\left( a \left|-\alpha\right\rangle +b \left| \alpha\right\rangle\right),\label{eq25}
\end{equation}
where $|a|^{2} + |b|^{2} =1$ and the normalization constant become
\begin{equation}
	N_{\alpha} = 1 + \exp\left[-2|\alpha|^2\right]\left(ab^{*}+a^{*}b\right),
\end{equation}
thus, the effect of a beam splitter on an input state consisting of a state (\ref{eq25}) in one mode and the vacuum state in the other is achieved as follows
\begin{equation}
	\left|\varPhi_{T} \right\rangle={\cal\hat{B}}\left(\theta\right)\left|\Psi_{\alpha}\right\rangle=\frac{1}{\sqrt{N_{\alpha}}}\left( a \left|-\alpha T	,-\alpha R\right\rangle +b \left| \alpha T, \alpha R\right\rangle\right).\label{eq27}
\end{equation}
The above state (\ref{eq27}) describes both the quantum field and the loss modes that constitute the environment. The final state after transmission is therefore derived by performing a partial trace on all modes of the environment as
\begin{equation}
	\hat{\rho} =\sum_{n=0}^{\infty} \left\langle n|\varPhi_{T}\right\rangle\left\langle \varPhi_{T}|n\right\rangle.
\end{equation}
A simplified calculation yields
\begin{align}
	\hat{\rho}=&\frac{1}{N_{\alpha}}\left[ \mid a\mid^{2}\left|-\alpha T \right\rangle\left\langle-\alpha T\right|+ \mid b\mid^{2}\left|\alpha T \right\rangle\left\langle\alpha T\right|\right]\notag\\&+\frac{q_{R\alpha}}{N_{\alpha}}\left[ab^{*} \left|-\alpha T \right\rangle\left\langle\alpha T\right|+ ba^{*}\left|\alpha T \right\rangle\left\langle-\alpha T\right|\right], 
\end{align}
with the quantity $q_{R\alpha}$ is
\begin{equation}
	q_{R\alpha}= \exp\left[-2R^{2}\mid\alpha\mid^{2}\right].
\end{equation}
Alternatively, we can simply verify that the last equation can be written in terms of the input state (\ref{eq25}) as follows
\begin{equation}
	\begin{split}
		\hat{\rho}=\frac{N_{\alpha T}}{2N_{\alpha}}\bigl[&\left(1+q_{R\alpha} \right)\left|\Psi_{\alpha T}\right\rangle \left\langle \Psi_{\alpha T}\right|+\\
		&\left(1-q_{R\alpha} \right)\hat{\Sigma}_{z}\left|\Psi_{\alpha T}\right\rangle \left\langle \Psi_{\alpha T}\right|\hat{\Sigma}_{z}\bigr],
	\end{split}
\end{equation}
where $\hat{\Sigma}_{z}$ is the Pauli operator acting on the state $\left|\Psi_{\alpha T}\right\rangle$ as
\begin{equation}
	\hat{\Sigma}_{z}\left|\Psi_{\alpha T}\right\rangle = \frac{1}{\sqrt{N(\alpha
			t)}} \left( a \left|\alpha T\right\rangle - b \left|\alpha  T\right\rangle \right).
\end{equation}
It should be emphasized here that the application of the $\hat{\Sigma}_{z}$-operator on the state $\left|\Psi_{\alpha T}\right\rangle$ induces a phase flip on the basis of the realized qubit. Therefore, the transmission of quantum information encrypted in coherent states is influenced by two types of noise; The first one concerns the reduction of the amplitude of the coherent state and the second type related to the phase flip modeled by the application of $\hat{\Sigma}_{z}$ on the state $\left|\Psi_{\alpha T}\right\rangle$.
\subsection{Generating Bell coherent-states superpositions by a beam splitter}
Alternatively, it is possible to construct experimentally the Bell-cat states using a simple scheme with a 50-50 beam splitter (i.e., $\theta=\pi/4$). This can be achieved by sending a cat state of the form $\left|\sqrt{2}\alpha\right\rangle+\left|-\sqrt{2}\alpha\right\rangle$ and the vacuum into the two input ports to obtain results in the output
\begin{equation}
	\left|\varPhi_{\alpha,\alpha}^{\rm Bell} \right\rangle=\frac{1}{\sqrt{N_{\alpha}^{\rm Bell}}}\left(\left|
	\alpha\right\rangle\left| \alpha\right\rangle  +  \left|
	-\alpha\right\rangle\left|-\alpha\right\rangle \right),\label{bell}
\end{equation}
where the normalization factor is defined by
\begin{equation}
	N_{\alpha}^{\rm Bell} = 2 \left( 1 + \exp\left[-4\vert \alpha \vert^2\right]\right).
\end{equation}
Indeed, the difficulty of generating Bell states comes down to the construction of a source of cat states that can be produced for example by sending a coherent state into a nonlinear medium by displaying the Kerr effect \cite{Yurke1986}. An alternative approach to produce these approximate states via photon counters and linear optical devices is reported in \cite{Ralph2003,Dakna1997,Song1990}. Furthermore, through the use of inefficient photon detectors and beam splitters, Lund and his colleagues \cite{Lund2004} proposed another approach to producing high amplitude cat states from squeezed single photon states. Experimentally, it is still difficult to produce cat states, especially ones with high amplitudes or mean photon counts. The experimental results that have been published and were produced with current technology are promising. Photon subtraction from squeezed vacuum resulted in superpositions of weak coherent states with opposite phase, resembling a Schrödinger cat state \cite{Ourjoumtsev2007}. Additionally, it was reported in \cite{Gerrits2010} that homodyne detection and photon number states were used as resources for the experimental creation of arbitrarily large squeezed cat states.\par

As shown in the subsection above, after passing the beam splitter on the Bell-cat states (\ref{bell}), the resultant density is
\begin{align}\label{density-main}
	\rho_{AB} = \frac{N_{\alpha T}^{\rm Bell}}{N_{\alpha}^{\rm Bell}} \bigg[& \frac
	{1}{2}\left(1+q_{R\alpha}\right)\left|\varPhi_{\alpha,\alpha T}^{\rm Bell} \right\rangle\left\langle \varPhi_{\alpha,\alpha T}^{\rm Bell}\right|
	+\notag\\& \frac {1}{2}(1-q_{R\alpha}) \hat{\Sigma}_{z}\left|\varPhi_{\alpha,\alpha T}^{\rm Bell} \right\rangle\left\langle \varPhi_{\alpha,\alpha T}^{\rm Bell}\right| \hat{\Sigma}_{z}\bigg].
\end{align}
To construct a qubit mapping, we insert a two-dimensional basis generated by the vectors $\left| u_{\alpha}\right\rangle$ and $\left| v_{\alpha} \right\rangle$ for the first mode $A$ as
\begin{equation}\label{qubitA}
	\left| \alpha \right\rangle= a_{\alpha} \left| u_{\alpha}\right\rangle +
	b_{\alpha}\left| v_{\alpha}\right\rangle, \hspace{0.5cm} \left|-\alpha\right\rangle=
	a_{\alpha} \left| u_{\alpha}\right\rangle - b_{\alpha}\left| v_{\alpha}\right\rangle,
\end{equation}
where
\begin{equation}
	\vert a_{\alpha} \vert^2 + \vert b_{\alpha} \vert^2 = 1,\hspace{1cm}
	\vert a_{\alpha} \vert^2 - \vert b_{\alpha} \vert^2 =\left\langle
	-\alpha \vert \alpha\right\rangle.
\end{equation}
Without loss of generality, we assume that $a_{\alpha}$ and $b_{\alpha}$ are reals such as
\begin{equation}
	a_{\alpha} = \frac{\sqrt{1 + p}}{\sqrt{2}},\hspace{1cm}{\rm and}\hspace{1cm} b_{\alpha}=\frac{\sqrt{1 - p}}{\sqrt{2}},
\end{equation}
with $p=\exp\left[-2\vert\alpha \vert^{2}\right]$. For the second mode $B$, a two dimensional basis is constructed by the vectors $\vert u_{\alpha T}\rangle$ and $\vert v_{\alpha T} \rangle$ as
\begin{align}
	&\left|\alpha T\right\rangle= a_{\alpha T}\left| u_{\alpha T}\right\rangle + b_{\alpha T}\left| v_{\alpha T}\right\rangle,\notag\\&\left|-\alpha T\right\rangle = a_{\alpha T} \left|u_{\alpha T}\right\rangle- b_{\alpha T}\left| v_{\alpha T}\right\rangle,
\end{align}
where
\begin{equation}
	a_{\alpha T} = \frac{\sqrt{1 + p^{T^2}}}{\sqrt{2}},\hspace{1cm}{\rm and}\hspace{1cm}b_{\alpha T}=\frac{\sqrt{1 - p^{T^2}}}{\sqrt{2}}.
\end{equation}
In the computational basis covered by the two-qubit product states
\begin{align}
	&|1\rangle = |u_{\alpha } \rangle_A
	\otimes |u_{\alpha T} \rangle_B, \hspace{1cm}\quad |2\rangle = | u_{\alpha}
	\rangle_A \otimes | v_{\alpha T}\rangle_B,\notag\\&
	|3\rangle =|v_{\alpha }\rangle_A \otimes |u_{\alpha T}\rangle_B, \hspace{1cm} |4\rangle=|v_{\alpha }\rangle_A \otimes|v_{\alpha T}\rangle_{B},
\end{align}
the resulting density matrix (\ref{density-main}) can be written as
\begin{eqnarray}
	\rho_{AB}=\frac{2}{N_{\alpha}^{\rm Bell}}\left(
	\begin{array}{cccc}
		\Lambda_{a} & 0 & 0 &
		{\cal M}_{ab}^{+}\\
		0 & \Lambda_{ab} & {\cal M}_{ab}^{-}  & 0 \\
		0 & {\cal M}_{ab}^{-} &
		\Lambda_{ba} 
		& 0 \\
		{\cal M}_{ab}^{+} & 0 & 0 &
		\Lambda_{b} 
	\end{array}
	\right),  \label{Xform11}
\end{eqnarray}
with the entries given by
\begin{align}
	&\Lambda_{a}=\left(1+q_{R\alpha}\right)a^2_{\alpha}a^2_{\alpha T},\hspace{1cm}\Lambda_{b}=\left(1+q_{R\alpha}\right)b^2_{\alpha }b^2_{\alpha T}\notag\\&\Lambda_{ab}=\left(1-q_{R\alpha}\right)a^2_{\alpha}b^2_{\alpha T}, \hspace{1cm}\Lambda_{ba}=\left(1-q_{R\alpha}\right)b^2_{\alpha}a^2_{\alpha T},\notag\\& {\cal M}_{ab}^{\pm}=\left(1\pm q_{R\alpha}\right)a_{\alpha }a_{\alpha T}b_{\alpha }b_{\alpha T}.
\end{align}

\section{Analytical expressions of relevant quantum criteria}\label{SecIII}
\subsection{Quantum Parameter Estimation}
In this section, we briefly recapitulate the basic literature about any quantum metrology task and that used later in our calculations. Mainly, quantum estimation theory aims at determining the ultimate precision of all unknown physical parameters encoded in quantum systems. It is focused on making high precision measurements of given parameters exploiting quantum resources \cite{Huelga1997,Hauke2016,McCormick2019}. Typically, a complete quantum metrological process consists of three steps: The first step of this protocol is to prepare the probe state, i.e., the input state $\rho$. The second part is the parameterization, i.e. the encoding of the information about the unknown parameter $\theta$ that can be realized by a unitary evolution, ${\cal U}_{\theta}=e^{-\theta {\cal H}}$, generated by the local Hamiltonian ${\cal H}={\cal H}_{A}\otimes\mathbb{1}_{B}$ to give the evolved state as $\rho_{\theta}={\cal U}_{\theta}\rho{\cal U}_{\theta}^{\dagger}$. Then, we apply the measurement of an appropriate observable $K$ on the output state $\rho_{\theta}$. Lastly, we employ the classical estimation theory which is well studied in classical statistics \cite{SlaouiD2022}.\par

In estimation theory, solving a parameter estimation problem comes down to finding an estimator $\hat{\theta}$, which can be considered as an application that gives us a set of measurement results in the parameter space. Besides, the determination of the ultimate accuracy depends on the quantum Fisher information which is essential to obtain the quantum Cramér-Rao bound \cite{Helstrom1969}. The optimal estimators are those that saturate this Cramer-Rao inequality:
\begin{equation}\label{crrao}
	\mathrm{Var}(\hat{\theta})\geq \frac{1}{n\mathcal{F}(\rho_{\theta})},
\end{equation}
where $n$ is the number of the measurements performed and the efficiency of an estimator (i.e., its intrinsic uncertainty) is quantified by the variance $\mathrm{Var}(\hat{\theta})$. Thus, it is clear that the inverse of the QFI $\mathcal{F}(\rho_{\theta})$ provides a lower bound on the statistical estimate of an unknown parameter. Indeed, the QFI is the maximal information about the estimated parameter $\theta$ that can be obtained from the optimal measurements \cite{Holevo1982,Liu2019}.\par
 
In terms of the symmetric logarithmic derivative (SLD) operator $\mathcal{L}_{\theta}$, which fulfills $2\partial_{\theta}\rho_{\theta}=\left\lbrace\rho_{\theta}, \mathcal{L}_{\theta} \right\rbrace$, the QFI appearing in the Cramér-Rao bound (\ref{crrao}) can be evaluated as $\mathcal{F}_{\theta}(\rho_{\theta})=\mathrm{Tr}\{\rho_{\theta} \mathcal{L}_{\theta}^2\}$, where the SLD operator $L_{\theta}$ is calculated from the equation
\begin{equation}
\mathcal{L}_{\theta}=2\int_{0}^{\infty}\exp\left[-\rho_{\theta}t \right]\frac{\partial\rho_{\theta}}{\partial\theta} \exp\left[-\rho_{\theta}t \right].
\end{equation}
For a diagonalized density matrix like $\rho_{\theta}=\sum_{i}\eta_{i}\left|\varphi_{i} \right\rangle\left\langle\varphi_{i}\right|$, it is simple to chech that the QFI of $\rho_{\theta}$ with respect to a Hermitian operator $\mathcal{H}$ takes the form 
\begin{align}
	\mathcal{F}_{\theta}(\rho_{\theta},\mathcal{H})=\sum_{i}\frac{{(\partial_{\theta}\eta_{i})^{2}}}{\eta_{i}}
	+\sum_{i\neq j}\frac{{2(\eta_{i}-\eta_{j})^{2}}}{\eta_{i}+\eta_{j}}
	\vert\langle\varphi_{i}\vert\partial_{\theta}\varphi_{j}\rangle\vert^{2},\label{formula:qfi}
\end{align}
where $\eta_{i}$ is the eigenvalue of the estimated state $\rho_{\theta}$, $\vert\varphi_{i}\rangle$ is the corresponding eigenvector, $\partial_{\theta}(\cdot)$ means the partial derivative, and the summation run over all eigenvalues satisfying $\eta_{i}\neq0$ and $\eta_{i}+\eta_{j}\neq0$. For non-full rank density matrix, the expression of the QFI $\mathcal{F}_{\theta}$ can be rewritten as \cite{Liu2013}
\begin{align}
	\mathcal{F}_\theta(\rho_\theta,\mathcal{H})=&\sum_{i=1}^r\frac{(\partial_\theta\eta_i)^2}{\eta_i}
	+\sum_{i=1}^r4\eta_i\langle\partial_\theta\varphi_i\vert\partial_\theta\varphi_i\rangle
	\notag\\&-\sum_{i,j=1}^r\frac{8\eta_i\eta_j}{\eta_i+\eta_j}\vert\langle\varphi_i\vert\partial_\theta\varphi_j\rangle\vert^2,
\end{align}
where $r$ is the rank of the density matrix. When $\rho_{\theta}$ is pure state, the QFI can be simplified as
\begin{equation}
\mathcal{F}_\theta(\vert\varphi\rangle)=4\left(\langle\partial_\theta\varphi\vert\partial_\theta\varphi\rangle
-\vert\langle\varphi\vert\partial_\theta\varphi\rangle\vert^2\right).
\end{equation}
\subsection{Quantum Correlation Quantifiers}
Quantification and characterization of quantum correlations are one of the challenges that have driven the development of quantum information processing in recent years. Given its fundamental importance and its increasingly practical nature, quantum correlation is a crucial resource for many aspects of quantum information theory \cite{Roga2016,Adesso2016}. In fact, interacting quantum systems exhibit various types of correlations, and among its various components is quantum entanglement \cite{Bengtsson2017}. These quantum correlations have been implemented by using different quantifiers and can be classified into two categories; The first class of measures is based on entropy functions, we quote for instance the entropic quantum discord and measurement-induced disturbance \cite{Vinjanampathy2012,Luo2008}. The second class concerns the geometric metrics quantified via different $p$-norm such as the Schatten one-norm, Hilbert-Schmidt norm and Bures norm \cite{Chruscinski2006,Planat2006,Basu2007}. We provide here the analytical expressions for the quantum correlation quantifiers that are related to the accuracy of the parameter estimates; namely, the entropic quantum discord based on von Neumann entropy, the local quantum Fisher information and the local quantum uncertainty that is based on the Skew information. We also establish the relationship between them.
\subsubsection{\bf Entropic Quantum Discord}
Basically, the quantum states can carry both classical and quantum correlations, however quantum entanglement quantifiers are incapable of detecting quantum correlations other than non-local ones. Contrary to the classical scenario, the measurement process in quantum mechanics disturbs the state in which a physical system is. This special property means that the disturbance induced by a measurement on a given quantum state is a good indication of its "quantumness". Inspired by this perturbation, the concept of entropic quantum discord (QD) has been introduced as a global quantifier of quantum correlations \cite{Henderson2001,Ollivier2002}, even those beyond entanglement.\par 

The concept of QD of a bipartite quantum system is defined as the difference between total and classical correlations. It is the measure of the amount of information that cannot be obtained by performing the measurement on a single subsystem. It can be mathematically quantified by the difference between the original quantum mutual information, ${\cal I}\left(\rho_{AB}\right)=S\left(\rho_{A}\right)+S\left(\rho_{B}\right)-S\left(\rho_{AB}\right)$, and the local measurement-induced quantum mutual information ${\cal CC}\left(\rho_{AB}\right)$, namely
\begin{equation}
{\cal QD}\left(\rho_{AB}\right)={\cal I}\left(\rho_{AB}\right)-{\cal CC}\left(\rho_{AB}\right),\label{eq17}
\end{equation}
with $S\left(\rho \right)=-{\rm Tr}\left(\rho\log_{2}\rho\right)$ is the von Neumann entropy and $\rho_{A}$, $\rho_{B}$ are the reduced density matrices of the composite system $\rho_{AB}$. In order to ensure that all classical correlations ${\cal CC}\left(\rho_{AB}\right)$ are accounted for, we need to maximize it over the set of positive operator valued measurements (POVM) $\left\lbrace \Pi_{k}^{B}\right\rbrace$ on subsystem B, which satisfy $\sum_{k}\Pi_{k}^{B^{\dagger}}\Pi_{k}^{B}=\mathbb{1}$, i.e.,
${\cal CC}\left(\rho_{AB}\right)={Sup}_{\left\lbrace \Pi_{k}^{B}\right\rbrace}{\cal I}\left(\rho_{AB}|\left\lbrace \Pi_{k}^{B}\right\rbrace \right)$. It reduced to
\begin{equation}
{\cal CC}\left(\rho_{AB}\right)=S\left(\rho_{A}\right)-\min_{\left\lbrace \Pi_{k}^{B}\right\rbrace}S\left(\rho_{B}|\left\lbrace \Pi_{k}^{B}\right\rbrace \right), \label{eq18}
\end{equation}
where the conditional entropy based on the measurement $\left\lbrace \Pi_{k}^{B}\right\rbrace$ is given by $S\left(\rho_{B}|\left\lbrace \Pi_{k}^{B}\right\rbrace \right)=\sum_{k}S\left(\rho_{k}^{B}\right)$, $p_{k}={\rm Tr}\left[\left(\mathbb{1}_{A}\otimes\Pi_{k}^{B} \right)\rho_{AB}\left(\mathbb{1}_{A}\otimes\Pi_{k}^{B}\right)\right]$ and $\rho_{k}^{B}=\left(\left(\mathbb{1}_{A}\otimes\Pi_{k}^{B} \right)\rho_{AB}\left(\mathbb{1}_{A}\otimes\Pi_{k}^{B}\right) \right)/p_{k}$ are the probability and the resulting state of subsystem $B$ with respect to the measurement outcome $k$. From Eqs.(\ref{eq17}) and (\ref{eq18}), QD can be redefined by the following expression
\begin{equation}\label{DAB}
	{\cal QD}\left(\rho_{AB}\right)= S\left(\rho_{A}\right)-S\left(\rho_{AB}\right)+\min_{\left\lbrace \Pi_{k}^{B}\right\rbrace}S\left(\rho_{B}|\left\lbrace \Pi_{k}^{B}\right\rbrace \right).
\end{equation}
In order to connect this quantifier to the entanglement measures, there is a close relationship between the entanglement of formation (EOF) and the QD, so-called the Koashi-Winter relation \cite{Koachi2004}:
\begin{equation}
{\cal QD}\left(\rho_{AB}\right)- S\left(\rho_{A}\right)+S\left(\rho_{AB}\right)={\cal E}_{F}\left(\rho_{BE}\right),\label{RKW}
\end{equation}
where a density matrix $\rho_{ABE}$ being a purification of $\rho_{AB}$, with $A$ and $B$ representing two subsystems and $E$ representing the environment. ${\cal E}_{F}\left(\rho_{BE}\right)$ is the EOF of $\rho_{BE}={\rm Tr}_{A}\left(\rho_{ABE}\right)$ and it is defined by the convex roof,
\begin{equation}
{\cal E}_{F}\left(\rho_{BE}\right)=\min_{\left\lbrace p_{i},\left| \psi_{i}\right\rangle_{BE}\right\rbrace}\sum_{i}{\cal E}_{F}\left(\left| \psi_{i}\right\rangle_{BE}\right),\label{EF}
\end{equation}
where the minimum is taken over all possible pure-state decompositions $\left\lbrace p_{i},\left| \psi_{i}\right\rangle_{BE}\right\rbrace$ of $\rho_{BE}$ and ${\cal E}_{F}\left(\rho_{BE}\right)$ represents the entanglement of formation for pure states $\left\lbrace\left| \psi_{i}\right\rangle_{BE}\right\rbrace$. It is given by the von Neumann entropy of the reduced subsystem $\rho_{B}$, i.e., ${\cal E}_{F}\left(\left| \psi_{i}\right\rangle_{BE}\right)=S\left(\rho_{B} \right)$ where $\rho_{B} ={\rm Tr}_{E}\left(\left| \psi_{i}\right\rangle_{BE}\left\langle\psi_{i}\right|\right)$. Indeed, the Koashi-Winter relation (\ref{RKW}) accounts for the trade-off between entanglement ${\cal E}_{F}\left(\rho_{BE}\right)$ (\ref{EF}) and classical correlations ${\cal CC}\left(\rho_{AB}\right)$ (\ref{eq18}), which means that the more the entanglement of the subsystem $\rho_{A}$ is shared with the environment $\rho_{E}$, the less classical information on $\rho_{A}$ is accessible via optimal measurements on the subsystem $\rho_{B}$. For an arbitrary two-qubit state $\rho_{BE}$, the entanglement of formation ${\cal E}_{F}\left(\rho_{BE}\right)$ is related to the Wootter's concurrence $\mathcal{ C}\left(\rho_{BE}\right)$ and can be expressed as
\begin{equation}\label{stild-min}
	{\cal E}_{F}\left(\rho_{BE}\right)=h\left( \frac{1+\sqrt{1-|\mathcal{ C}\left(\rho_{BE}\right)|^{2}}}{2}\right) 
\end{equation}
where $h(x)=-x \rm log_{2}x-(1-x)\rm log_{2}(1-x)$ is the binary entropy function. At this point, we invoke the explicit formula of concurrence,
\begin{equation}
	\mathcal{ C}\left(\rho_{BE}\right)=\max\left\lbrace0,\sqrt{\vartheta_{1}}-\sqrt{\vartheta_{2}}-\sqrt{\vartheta_{3}}-\sqrt{\vartheta_{4}} \right\rbrace, 
\end{equation}
by computing analytically the eigenvalues $\vartheta_{i}$ of the operator $\rho_{BE}\tilde{\rho}_{BE}$ in descending order, $\vartheta_{i}\geq\vartheta_{i+1}$, where $\tilde{\rho}_{BE}=\left(\sigma_{B}^{y}\otimes\sigma_{E}^{y}\right)\rho_{BE}^{*}\left(\sigma_{B}^{y}\otimes\sigma_{E}^{y}\right)$ is the "spin-flipped" density matrix.
\subsubsection{\bf Geometric Quantum Discord via the Schatten $1$-norm}

As evaluating the entropic quantum discord involves an optimization procedure, the analytical results are only known for a few families of two-qubit states. To overcome this problem, a geometric version of entropy discord has been defined using the fact that entropy quantum discord cancels for classically correlated states \cite{BellomoFranco2012,BellomoGiorgi2012}. This is completely analogous to geometric measures of entanglement which are defined in terms of distances from the set of separable states \cite{Hubener2009}. Along these lines, a necessary and sufficient condition for the existence of non-zero quantum discord has been achieved and a geometric method of quantizing quantum discord has been proposed \cite{Spehner2013,Luo2010,Jin2012}.\par

Geometric quantum discord quantifies the amount of quantum correlation existing in a quantum state by employing the Schatten distance between the state of interest $\rho_{AB}$ and its nearest classically correlated state $\chi$ as \cite{Paula2013}
\begin{equation}
{\cal Q}_{g}\left(\rho_{AB}\right)=\min_{\chi\in\Omega_{0}}\left(\parallel\rho_{AB}-\chi\parallel_{p} \right)^{p},\label{dg}
\end{equation}
where $\parallel Z\parallel_{p}=\left[{\rm Tr}\left(Z^{\dagger}Z\right)^{\frac{p}{2}}\right]^{\frac{1}{p}}$ denotes the Schatten p-norms, which reduces to the trace norm for $p=1$ and to the Hilbert Schmidt norm for $p=2$. Using equation (\ref{dg}) for $p=1$, the geometric quantum discord, also called the trace quantum discord, is reduced to
\begin{equation}
{\cal Q}_{gT}\left(\rho_{AB}\right)=\min_{\chi\in\Omega_{0}}\parallel\rho_{AB}-\chi\parallel_{1},
\end{equation}
with $\parallel\rho_{AB}-\chi\parallel_{1}={\rm Tr}\left[\left(\rho_{AB}-\chi \right)^{\dagger}\left(\rho_{AB}-\chi\right)\right]$ and $\chi=\sum_{k}p_{k}\Pi_{k,A}\otimes\rho_{k,B}$. The minimization method on the set of classical states for two-qubit $X$-states with one norm was proposed in ref.\cite{Ciccarello2014}, and the analytical expression of the trace-norm geometric discord was found as
\begin{equation}
{\cal Q}_{gT}\left(\rho_{AB}\right)=\sqrt{\frac{{\cal R}_{11}^{2}{\cal R}_{\max}^{2}-{\cal R}_{22}^{2}{\cal R}_{\min}^{2}}{{\cal R}_{\max}^{2}-{\cal R}_{\min}^{2}+{\cal R}_{11}^{2}-{\cal R}_{22}^{2}}},\label{GTD}
\end{equation}
where ${\cal R}_{\alpha\beta}={\rm Tr}\rho_{AB}\left({{\sigma_\alpha}\otimes{\sigma_\beta}}\right)$ are the correlation matrix elements of the density matrix $\rho_{AB}$ with ${\cal R}_{\min}^{2}=\min\left\lbrace{\cal R}_{11}^{2},{\cal R}_{33}^{2} \right\rbrace$ and ${\cal R}_{\max}^{2}=\max\left\lbrace{\cal R}_{33}^{2},{\cal R}_{22}^{2}+{\cal R}_{30}^{2} \right\rbrace$.

\subsubsection{\bf Local Quantum Fisher Information}
According to an operational framework generally referred to as black box interferometry \cite{Girolami2014,Kim2018}, the ability to estimate the local phase shift applied to one of the entrances of an interferometer is measured by the local quantum Fisher information (LQFI). It quantifies the accuracy that such a bipartite quantum state allows for the estimation of a parameter, embedded in a unitary dynamics applied to a single subsystem, by employing the quantum Fisher information. In order to understand the role of quantum correlations in the phase estimation protocol, the LQFI is a bona fide measure of non-classical correlations in this scenario. It is defined by the minimum of the quantum Fisher information \cite{Kim2018}, namely
\begin{equation}\label{powerDEF}
	\mathcal{Q}_{\cal F}\left(\rho_{AB}\right)=\min_{H_{A}}\mathcal{F}_{\theta}\left(\rho_{AB}, {\cal H}_{A}\right),
\end{equation}
where the minimization is predicted over all possible choices of the local dynamics generated by an Hamiltonian ${\cal H}_{A}$ with a given spectrum. Practically, the probe states $\rho_{AB}$ with higher LQFI embody more reliable resources for quantum metrology, as they ensure lower variance in the estimate of $\theta$. The LQFI satisfies all the known criteria for a discord-like quantifier; it is non-negative and vanishes if and only if $\rho_{AB}$ is a classical state. It does not increase under local operations on the second qubit and satisfies the inequality $\mathcal{Q}_{\cal F}\left(\sum_{i}\eta_{i}\rho_{i},{\cal H}_{A}\right)\leq\sum_{i}\eta_{i}\mathcal{Q}_{\cal F}\left(\rho_{i},{\cal H}_{A}\right)$. It is also invariant under local unitary transformations and reduces to an entanglement monotone for pure states.\par

In the relevant case where the subsystem $A$ is a qubit, the general form of such local Hamiltonian's reduced to ${\cal H}_{A}=\vec{r}.\vec{\sigma}=\sum_{i=1}^{3}r_{i}\sigma_{i}$, where $\vec{\sigma}=\left\lbrace\sigma_{1},\sigma_{2},\sigma_{3}\right\rbrace$ is the vector of the Pauli matrices, $\vec{r} =\{\sin\varphi\:\cos\phi, \sin\varphi\:\sin\phi,\cos\varphi\}$ and $|\vec{r}|=1$. Explicitly, this operational quantifier is straightforward to compute analytically for an arbitrary $2\otimes d$ quantum system and its closed analytical formula is given by
\begin{equation}
	\mathcal{Q}_{\cal F}\left(\rho_{AB}\right)=1-\zeta_{\max}\left[{\cal M}_{ij}\right],\label{QF}
\end{equation}
with $\zeta_{\max}$ denotes the smallest eigenvalue of the symmetric matrix ${\cal M}$ whose matrix elements are defined as below:
\begin{equation}
	{\cal M}_{ij}=\sum_{k\neq l}\frac{2\eta_{k}\eta_{l}}{\eta_{k}+\eta_{l}}\langle\varphi_{l}|\sigma_{i}\otimes \openone_{B}|\varphi_{k}\rangle\langle\varphi_{k}|\sigma_{j}\otimes \openone_{B}|\varphi_{l}\rangle,\label{Mij}
\end{equation}
where the summation is performed under the condition $\eta_{k}+\eta_{l}>0$. The $\mathcal{Q}_{\cal F}=1$ for maximally correlated Bell cat states and it is zero value for classically correlated states. Otherwise, in the interval $0<\mathcal{Q}_{\cal F}<1$, the LQFI reveals the collective correlations between the cat states.
\subsubsection{\bf Local Quantum Uncertainty based on Weak Measurements}
The local quantum uncertainty (LQU), as introduced by Girolami and his collaborators \cite{Girolami2013}, is a significant figure of merit for quantum correlations that go beyond those described by entanglement. Despite the popularity and convenience of QD, the notorious difficulty of its calculation poses a curious and irritating problem. A particular merit of LQU is that it can be evaluated analytically for any ($2\times d$)-dimensional quantum state \cite{Slaoui2018,Slaoui2019}. This quantity is constitutes an alternative tool to evaluate the analytical expressions of quantum correlations encompassed in any bipartite systems. This measurement satisfies all the known criteria for a discord-like quantifier for general mixed states and also deeply related to quantum Fisher information in the context of quantum metrology \cite{Luo2003}.
The LQU is defined as the minimum skew information \cite{Wigner1963} achievable by a single local measurement
\begin{equation}
	\mathcal{U}\left(\rho_{AB}\right)= \min_{K_\Lambda} \mathcal{I}(\rho_{AB}, K_{\Lambda}^{A} \otimes
	\openone_B), \label{LQU}
\end{equation}
where $K_\Lambda=K_{\Lambda}^{A}\otimes\openone_{B}$ is a local observable on subsystem $A$ and $\openone_{B}$ is the identity operator acting on the qubit $B$. To find the optimization observables that minimize the skew information in equation (\ref{LQU}), note that $K_{\Lambda}^{A}$ is a Hermitian operator with a non-degenerate spectrum and can be parameterized as $K_{\Lambda}^{A}=V_{A}{\rm diag}\left(\Lambda\right)V_{A}^{\dagger}$, where $V_{A}$ is varied over the special unit group of the subsystem $A$. Moreover notice that, the skew information $\mathcal{I}(\rho_{AB},K_{\Lambda}^{A} \otimes \openone_B)$ reflects the non commutation between the quantum state $\rho$ and the observable $K_{\Lambda}^{A}$, which is defined as
\begin{equation}
	\mathcal{I}(\rho_{AB},K_A \otimes
	\openone_B)=-\frac{1}{2}{\rm
		Tr}([\sqrt{\rho_{AB}}, K_A \otimes
	\openone_B]^{2}).
\end{equation}
The minimization in equation (\ref{LQU}) can be done exactly for the bipartite $2\otimes d$ systems and the LQU can be analytically calculated as
\begin{equation}
	\mathcal{U}\left(\rho_{AB}\right)= 1 - \max\{\lambda_1, \lambda_2, \lambda_3\},  \label{LQU1}
\end{equation}
where $\lambda_{i}$'s are the eigenvalues of the $3\times3$ symmetric matrix $W$ whose matrix elements are defined by,
\begin{equation}\label{w-elements}
	\omega_{ij} \equiv  {\rm
		Tr}\{\sqrt{\rho_{AB}}(\sigma_{i}\otimes
	\openone_B)\sqrt{\rho_{AB}}(\sigma_{j}\otimes \openone_B)\},
\end{equation}
with $i,j = 1, 2, 3$. For a pure quantum state, the LQU reduces to the linear entropy of entanglement $\mathcal{U}\left(\left|\psi^{AB} \right\rangle \left\langle \psi^{AB} \right|\right)=2\left[1-{\rm Tr\left(\rho_{A} \right)^{2}} \right]$. Besides, the perturbation of the measured state was much stronger due to the fact that the information obtained was small when using the von Neumann projection. However, the information of the quantum states was recoverable after a weak measurement in order to minimize the impact on the initial quantum state, which induces a relatively weak modification on the measured state \cite{Aharonov1988,Tamir2013,Story1991}. Here, we examined what happens with the local quantum uncertainty when the projective measurements are replaced by weak-measurement dichotomic operators.\par

As introduced by Brun and Oreshkov \cite{Oreshkov2005}, the weak measurements can be formulated by using the pre- and post-selected quantum systems \cite{Aharonov1988} along with the projective measurement operator formalism. Thus, a quantum measurement with any number of outcomes can be constructed as a sequence of measurements with two outcomes. In such a formalism, the weak measurement operators are given by
\begin{equation}
	P\left(\chi\right)=\sqrt{\frac{1-\tanh\left(\chi \right)}{2}}\Pi_{0} +\sqrt{\frac{1+\tanh\left(\chi \right)}{2}}\Pi_{1},
\end{equation}
and
\begin{equation}
	P\left(-\chi\right)=\sqrt{\frac{1+\tanh\left(\chi \right)}{2}}\Pi_{0} +\sqrt{\frac{1-\tanh\left(\chi \right)}{2}}\Pi_{1},
\end{equation}
where $\Pi_{0}$ and $\Pi_{1}$ are two orthogonal projectors whose sum $\Pi_{0}+\Pi_{1}=\openone$ is the identity, $P\left(\chi\right)$ and $P\left(-\chi\right)$ describe a measurement with $\chi\in{\cal R}$ is the measurement strength parameter. If $\chi=\epsilon$, where $\mid\epsilon\mid<<1$, the measurement is weak. When $\chi=0$, i.e., $P\left(0\right)=1/\sqrt{2}$, the weak measurement causing no change from the initial state. It is obvious that $P^{\dagger}\left(\chi\right)P\left(\chi\right)+P^{\dagger}\left(-\chi\right)P\left(-\chi\right)=1$ and $\left[P\left(\chi\right),P\left(-\chi\right)\right]=0$. In addition, $\lim_{\chi\longmapsto\infty}P\left(-\chi\right)=\Pi_{0}$ and $\lim_{\chi\longmapsto\infty}P\left(\chi\right)=\Pi_{1}$. Using the compact expression of the weak measurement operators, 
\begin{equation}
P\left(\pm\chi\right)=\alpha\left(\pm\chi\right)\Pi_{0} +\alpha\left(\mp\chi\right)\Pi_{1},
\end{equation}
where $\alpha\left(\pm\chi\right)=\sqrt{\frac{1\mp\tanh\left(\chi\right)}{2}}$, the action of the weak operators can be mapped as
\begin{equation}
P\left(\pm\chi\right)=\frac{1}{2}\left[\alpha\left(+\chi\right)+\alpha\left(-\chi\right)\right]\openone_{A}\pm \frac{1}{2}\left[\alpha\left(+\chi\right)-\alpha\left(-\chi\right)\right]\sigma_{Z}.
\end{equation}
Within this representation, a particular local observable $P\left(\pm\chi\right)$ with non-degenerate spectrum $\left\lbrace\alpha\left(\chi\right),\alpha\left(-\chi\right) \right\rbrace$ can be parameterized as
\begin{align}
	P\left(\pm\chi\right)&=U_{A}\left( \frac{\alpha\left(+\chi\right)+\alpha\left(-\chi\right)}{2}\openone_{A}\pm \frac{\alpha\left(+\chi\right)-\alpha\left(-\chi\right)}{2}\sigma_{Z}\right)U_{A}^{\dagger}\notag\\& =\frac{\alpha\left(+\chi\right)+\alpha\left(-\chi\right)}{2}\openone_{A}\pm \frac{\alpha\left(+\chi\right)-\alpha\left(-\chi\right)}{2}\vec{n}.\vec{\sigma}^{A}\notag\\&=\beta_{+}\openone_{A}\pm\beta_{-}\vec{n}.\vec{\sigma}^{A},
\end{align}
where $\vec{n}$ is a unit vector and $\beta_{\pm}=\left(\alpha\left(+\chi\right)\pm\alpha\left(-\chi\right)\right)/2$. After a straightforward calculation, the skew information obtained from the weak measurements can be written as follows
\begin{align}
\mathcal{I}_{W}\left(\rho_{AB},P\left(\pm\chi\right)\right)&=-\frac{1}{2}{\rm Tr}\left\lbrace \beta_{+}\left[\sqrt{\rho_{AB}},\openone_{A}\right]
+\beta_{-}\left[\sqrt{\rho_{AB}},\vec{n}.\vec{\sigma}_{A}\right]\right\rbrace^{2}\notag\\&=-\frac{\beta_{-}^{2}}{2}{\rm Tr}\left\lbrace\left[\sqrt{\rho_{AB}},\vec{n}.\vec{\sigma}_{A}\right]\right\rbrace^{2}\notag\\&=\beta_{-}^{2}\mathcal{I}\left(\rho_{AB},\vec{n}.\vec{\sigma}_{A}\right).
\end{align}
It then follows that the local quantum uncertainty based on weak measurements taken the form
\begin{align}
&\mathcal{U}_{W}\left(\rho_{AB}\right)=\min_{K_A}\mathcal{I}_{W}\left(\rho_{AB},P\left(\pm\chi\right)\right)=\beta_{-}^{2}\min_{n.\sigma}\mathcal{I}\left(\rho_{AB},\vec{n}.\vec{\sigma}_{A}\right)\notag\\&=\beta_{-}^{2}\min_{n.\sigma}\left\lbrace1-{\rm Tr}\left(\sqrt{\rho_{AB}}\left(n_{i}\sigma_{i}\otimes\openone_{B} \right)\sqrt{\rho_{AB}}\left(n_{j}\sigma_{j} \otimes\openone_{B}\right)  \right)  \right\rbrace\notag\\&=\frac{\cosh\left(\chi\right)-1}{4\cosh\left(\chi\right)}\left[1 - \max\{\lambda_1, \lambda_2, \lambda_3\}\right].\label{UW}
\end{align}
Obviously, one can easily evaluate the local quantum uncertainty based on weak measurements for qubit-qudit quantum systems provided one has the elements of the $3\times3$ symmetric matrix $W$ (\ref{w-elements}). Further, we can easily show that this new measure is a faithful measure of quantum correlations like the local quantum uncertainty captured by the strong measurement; it is vanishes for classically correlated states, reduced to the linear entropy of entanglement for pure states and invariant under local quantum operations acting on the unmeasured qubit. For the two-qubit quantum states whose density matrices are $X$-shaped, which are of interest here and which are used in several quantum information problems, it is straightforward to analytically evaluate the weak measurement-induced local quantum uncertainty via the computation of the total correlation tensor occurring in the Fano-Bloch decomposition.\par
In the Fano Bloch representation, the density matrix $\rho_{AB}$ can be written as
\begin{equation}
	\rho_{AB}=\frac{1}{4}\sum_{\alpha \beta }{\cal R}_{\alpha \beta }{\sigma _\alpha } \otimes {\sigma _\beta },
\end{equation}
and for the $X$-type states, the matrix elements of Eq.(\ref{w-elements}) are given by
\begin{eqnarray}
{w_{11}} &=& \left( {\sqrt {{\eta _1}}  + \sqrt {{\eta_4}} } \right)\left( {\sqrt {{\eta_2}}  + \sqrt {{\eta_3}} } \right)   \nonumber \\
&+&{\frac{{ {\cal R}_{11}^2 - {\cal R}_{22}^2 + {\cal R}_{12}^2 - {\cal R}_{21}^2 +{\cal R}_{03}^2 - {\cal R}_{30}^2}}{{4\left( {\sqrt {{\eta _1}}  + \sqrt {{\eta _4}} } \right)\left( {\sqrt {{\eta_2}}  + \sqrt {{\eta _3}} } \right)}}},
\end{eqnarray}
\begin{eqnarray}
{w_{22}} &=& \left( {\sqrt {{\eta_1}}  + \sqrt {{\eta_4}} } \right)\left( {\sqrt {{\eta_2}}  + \sqrt {{\eta_3}} } \right) \nonumber \\
&+&\frac{{ {\cal R}_{22}^2 - {\cal R}_{11}^2 + {\cal R}_{21}^2 - {\cal R}_{12}^2  + {\cal R}_{30}^2 - {\cal R}_{03}^2}}{{4\left( {\sqrt {{\eta _1}}  + \sqrt {{\eta _4}} } \right)\left( {\sqrt {{\eta_2}}  + \sqrt {{\eta_3}} } \right)}},
\end{eqnarray}
\begin{eqnarray}
{w_{33}} &=& \frac{1}{2}\left( 1+2\sqrt {\eta_{1}\eta_{4}}  + 2\sqrt {\eta_{2}\eta_{3}}  \right) \nonumber \\
&+&\frac{{{{\left( {{{\cal R}_{30}} + {{\cal R}_{03}}} \right)}^2} - {{\left( {{{\cal R}_{11}} - {{\cal R}_{22}}} \right)}^2} - {{\left( {{{\cal R}_{12}} + {{\cal R}_{21}}} \right)}^2}}}{{8{{\left( {\sqrt {{\eta _1}}  + \sqrt {{\eta_4}} } \right)}^2}}} \nonumber  \\
&+&\frac{{{{\left( {{{\cal R}_{03}} - {{\cal R}_{30}}} \right)}^2} - {{\left( {{{\cal R}_{11}} + {{\cal R}_{22}}} \right)}^2} - {{\left( {{{\cal R}_{12}} - {{\cal R}_{21}}} \right)}^2}}}{{8{{\left( {\sqrt {{\eta_2}}  + \sqrt {{\eta_3}} } \right)}^2}}},
\end{eqnarray}
\begin{eqnarray}
	{w_{12}} &=& {w_{21}}  = \frac{1}{2}\frac{{{{\cal R}_{11}}{{\cal R}_{21}} + {{\cal R}_{22}}{{\cal R}_{12}}}}{{\left( {\sqrt {{\eta_1}}  + \sqrt {{\eta_4}} } \right)\left( {\sqrt {{\eta_2}}  + \sqrt {{\eta_3}} } \right)}},
\end{eqnarray}
and
\begin{eqnarray}
{w_{13}} &=& {w_{31}} = {w_{23}} = {w_{32}} = 0,
\end{eqnarray}
where $\eta_i (i = 1, 2, 3, 4)$ are the eigenvalues of the density matrix $\rho_{AB}$. Hereafter, we discuss the dynamics of these various quantum quantifiers on our theoretical model characterized by the density matrix (\ref{Xform11}).
\subsection{Measure of Quantum Coherence}
Before discussing the proper measure of quantum coherence applied to our system, we briefly review the framework of coherence measures by focusing on a general d-dimensional Hilbert space ${\cal H}$ with reference orthonormal basis $\left\lbrace\left| i\right\rangle \right\rbrace_{i=1,2,..,d}$. When the quantum state is diagonal in this local reference basis, the state is called incoherent (classical state) and takes the form $\delta=\sum_{i=1}^{d}p_{i}\left|i \right\rangle\left\langle i\right|$, with $\delta\in I$ where $I$ is a set of incoherent states (i.e., $\delta$ has no superposition) \cite{Aberg2014}. Every quantum state that cannot be described in this form is considered as a coherent state, implying that the quantum coherence is basis-dependent.\par

Similar to quantum entanglement theory which establishes the sets of local operations and classical communication and separable states, a rigorous framework has been developed for the quantification of quantum coherence based on the concepts of inconsistent operations and incoherent states \cite{Baumgratz2014}. Such incoherent operations are the completely positive linear trace preserving map which maps an incoherent state to an incoherent state and no coherence generation could be observed. Meanwhile, it points out that a good coherence measure from the resource theoretic perspective of quantum coherence should satisfy these series of (axiomatic) necessary conditions: ($i$) Incoherent states should not have any coherence, ($ii$) Incoherent operations cannot increase the quantum coherence as well as the average coherence is not increased under selective measurements (monotonicity), and finally, ($iii$) the quantum coherence is not increased under the mixing of quantum states (convexity) \cite{Baumgratz2014}. Several figures of merit for quantum coherence have been introduced and assessed, including the relative entropy of coherence (coherence cost) \cite{Bu2017}, distance-based coherence \cite{Streltsov2015}, coherence of formation \cite{Winter2016}, ${\rm lp}$-norm coherence \cite{Streltsov2015}, robustness of coherence \cite{Napoli2016}, coherence via fidelity \cite{Liu2017}, via quantum skew information \cite{Yu2017}, and via Tsallis relative entropy \cite{Rastegin2016}.\par

Interestingly, we can classify all the above coherence measures into two categories depending on whether the measure is based on the entropy functional or has a metric character that involves a geometric structure. In the current work, we use a coherence measure based on the quantum version of the Jensen-Shannon divergence, which has several mathematically favorable properties with metric properties and an entropic nature \cite{Radhakrishnan2016}. To this end, the quantum Jensen-Shannon divergence, denoted by ${\cal J}\left(\rho,\delta\right)$, is defined in Ref.\cite{Lin1991}:
\begin{equation}
{\cal J}\left(\rho,\delta\right)=\frac{1}{2}\left[S\left(\rho\parallel\frac{\rho+\delta}{2}\right)+S\left(\delta\parallel\frac{\rho+\delta}{2}\right)\right],\label{JDivergence}
\end{equation}
where the function $S\left(\rho\parallel\delta\right)$ is the relative entropy defined as
\begin{equation}
S\left(\rho\parallel\delta\right)={\rm Tr}\left(\rho\log_{2}\rho-\rho\log_{2}\delta \right). 
\end{equation}
In terms of von Neumann entropy, the above quantum Jensen–Shannon divergence (\ref{JDivergence}) can be written as
\begin{equation}
{\cal J}\left(\rho,\delta\right)=S\left(\frac{\rho+\delta}{2}\right)-\frac{1}{2}\left[S\left(\rho\right)+S\left(\sigma \right)\right].
\end{equation}
Although the square root of quantum Jensen–Shannon divergence obeys the distance axioms as well as the qualifies as a metric, hence it is the figure of merit for quantum coherence \cite{Radhakrishnan2016}. Towards this end, we have
\begin{align}
{\cal QC}\left(\rho\right)&=\min_{\delta\in I}\sqrt{{\cal J}\left(\rho,\delta\right)}\notag\\&=\sqrt{S\left(\frac{\rho+\rho_{d}}{2}\right)-\frac{1}{2}\left[S\left(\rho\right)+S\left(\rho_{d}\right)\right]},\label{QC}
\end{align}
where $\rho_{d}$ is the diagonal part of quantum state $\rho$ (the closest incoherent state).
\section{Dynamics of quantum criteria in generated Bell coherent-states superpositions}\label{SecIV}
In order to investigate the dynamical evolution of the mentioned pairwise quantum criteria and their roles in improving the estimation accuracy of an unknown phase shift in quantum metrology, we have computed the quantities relevant for our purposes namely: concurrence entanglement, entropy quantum discord, quantum coherence, trace norm geometric discord, local quantum Fisher information and local quantum uncertainty based on weak measurements, for Bell cat-states under amplitude damping.\par

By performing some obvious algebra and using the implicit form of the density matrix in equation (\ref{Xform11}), it is easy to check that the concurrence which quantifies the amount of entanglement in our system is given by
\begin{equation}
\mathcal{ C}\left(\rho_{AB}\right)=\frac{1}{2\left(1+\exp\left(-4\mid\alpha\mid^{2} \right) \right)}\max\left\lbrace0,{\hat{\mathcal{ C}}}_{1},{\hat{\mathcal{ C}}}_{2}\right\rbrace 
\end{equation}
with
\begin{align}
{\hat{\mathcal{ C}}}_{1}=\frac{1}{2}&\left[ \left(1+\exp\left(-2R^{2}\mid\alpha\mid^{2}\right)\right) \left(1+p \right)\left(1+p^{1-R^{2}} \right) -\right.\notag\\&\left. \left(1-\exp\left(-2R^{2}\mid\alpha\mid^{2}\right)\right)\sqrt{\left(1-p^{2} \right)\left(1-p^{2\left(1-R^{2}\right)} \right)}\right],
\end{align}
and
\begin{equation}
{\hat{\mathcal{ C}}}_{2}=\exp\left(-2R^{2}\mid\alpha\mid^{2}\right)\left(p^{2}-1\right)\left( 1-p^{2\left(1-R^{2}\right)}\right).
\end{equation}
On the other side, to obtain the explicit expression of the entropy quantum discord, we first compute the quantum mutual information that quantizes the total quantum correlations exhibited in our model. From (\ref{eq17}), the quantum mutual information is simply given by
\begin{align}
{\cal I}\left(\rho_{AB}\right)=&h\left(\frac{\left(1+p\right)^{2} }{2\left(1+p^{2} \right)}\right)+h\left(\frac{\left(1+p^{1-R^{2}} \right)\left(1+p^{1+R^{2}} \right) }{2\left(1+p^{2} \right)} \right)\notag\\&-h\left(\frac{\left(1+p^{R^{2}} \right)\left(1+p^{2-R^{2}} \right) }{2\left(1+p^{2} \right)} \right),
\end{align}
and therefore, the entropy quantum discord can be simply written as
\begin{align}
{\cal QD}\left(\rho_{AB}\right)&=h\left(\frac{1}{2}+\frac{\sqrt{1+p^{2}+p^{2\left(R^{2}+1 \right) }+p^{2\left(2-R^{2} \right)}}}{2\left(1+p^{2} \right)} \right)\notag\\&+h\left(\frac{\left(1+p\right)^{2} }{2\left(1+p^{2} \right)}\right)-h\left(\frac{1}{2}+\frac{p^{R^{2}}+p^{2-R^{2}}}{2\left(1+p^{2} \right)}\right),  
\end{align}
where $h\left(y\right)=-y\log_{2}y-\left(1-y \right)\log_{2}\left(1-y \right)$
represents the binary entropy. Subsequently, we investigated the analytic evolution of the geometric quantum discord based on the trace norm for Bell cat states under amplitude damping channel. In the Fano-Bloch representation of the density matrix (\ref{eq17}), the nonvanishing correlation matrix elements are given by:
\begin{equation*}
	{\cal R}_{11}=\frac{\sqrt{\left( 1-p^{2}\right)\left(1-p^{2\left(1-R^{2} \right)} \right)}}{1+p^{2}},\hspace{0.5cm}{\cal R}_{03}=\frac{p^{1-R^{2}}+p^{1+R^{2}}}{1+p^{2}},
\end{equation*}
and
\begin{equation*}
	{\cal R}_{22}=-p^{R^{2}}{\cal R}_{11},\hspace{0.5cm}{\cal R}_{33}=\frac{p^{R^{2}}+p^{2-R^{2}}}{1+p^{2}},\hspace{0.5cm} {\cal R}_{30}=\frac{2p}{1+p^{2}}.
\end{equation*}
Based on the above formalism, the geometric quantum discord via the Schatten $1$-norm (\ref{GTD}) can be evaluated to be
\begin{equation}
{\cal Q}_{gT}\left(\rho_{AB}\right)=\frac{1}{2}\sqrt{\frac{{\cal R}_{11}^{2}\left( {\cal R}_{\max}^{2}-p^{2R^{2}}{\cal R}_{\min}^{2}\right) }{{\cal R}_{\max}^{2}-{\cal R}_{\min}^{2}+{\cal R}_{11}^{2}\left( 1-p^{2R^{2}}\right) }},
\end{equation}
with
\begin{align}
{\cal R}_{\max}^{2}=\frac{1}{\left(1+p^{2}\right)^{2} }&\max\left\lbrace\left(p^{R^{2}}+p^{2-R^{2}} \right)^{2},4p^{2}+\right. \notag \\&\left. p^{2R^{2}} \left( 1-p^{2}\right)\left(1-p^{2\left(1-R^{2} \right) } \right)\right\rbrace,
\end{align}
and
\begin{equation}
{\cal R}_{\min}^{2}=\min\left\lbrace\frac{\left(1-p^{2} \right)\left(1-p^{2-2R^{2}} \right)}{\left(1+p^{2}\right)^{2} },\frac{\left(p^{R^{2}}+p^{2-R^{2}} \right)^{2}}{\left(1+p^{2}\right)^{2}}\right\rbrace. 
\end{equation}
Before analyzing the impact of beam splitting reflection coefficient and coherent state overlapping on entanglement and quantum correlation evolutions, it is convenient to find the amount of quantum coherence existing in the system which can be readily determined using the concept of Jensen-Shannon quantum divergence. After some algebraic manipulation and applying equation (\ref{JDivergence}), the quantum Jensen-Shannon divergence is found to be
\begin{align}
&{\cal J}\left(\rho,\rho_{d}\right)=-\sum_{\pm}{\cal L}_{\pm}\log_{2}{\cal L}_{\pm}-\sum_{\pm}\varGamma_{\pm}\log_{2}\varGamma_{\pm}+\notag\\&\frac{1}{2}\sum_{\pm}\gamma_{\pm}\log_{2}\gamma_{\pm}+\frac{1}{2}\left[\beta_{+++}\log_{2}\beta_{+++}+\beta_{-+-}\log_{2}\beta_{-+-}\right. \notag \\&\left. +\beta_{--+}\log_{2}\beta_{--+}+\beta_{+--}\log_{2}\beta_{+--} \right], 
\end{align}
with the quantities ${\cal L}_{\pm}$, $\varGamma_{\pm}$, $\gamma_{\pm}$ and $\beta_{\pm\pm\pm}$ are
\begin{align}
{\cal L}_{\pm}=\frac{2\left(1-p^{R^{2}} \right)\left(p^{R^{2}}-p^{2} \right)\pm\sqrt{\zeta}  }{8p^{R^{2}}\left(1+p^{2} \right) },
\end{align}
\begin{align}
	\varGamma_{\pm}=\frac{2\left(1+p^{R^{2}} \right)\left(p^{R^{2}}+p^{2} \right)\pm\sqrt{\Omega}  }{8p^{R^{2}}\left(1+p^{2} \right)},
\end{align}
\begin{equation}
	\gamma_{\pm}=\frac{\left(1\pm p^{R^{2}} \right)\left(1\pm p^{2+R^{2}} \right)  }{2\left(1+p^{2} \right) },
\end{equation}
\begin{align}
\beta_{\pm\pm\pm}=\frac{\left( 1\pm p\right)\left( 1\pm p^{1-R^{2}}\right)\left( 1\pm p^{R^{2}}\right) }{4\left(1+p^{2}\right)},
\end{align}
where
\begin{align}
\zeta=&p^{2}\left(3+p^{2} \right)-2p^{4+R^{2}}+p^{2R^{2}}\left(1+p^{2R^{2}}-2p^{R^{2}} \right)\notag\\&-14p^{2+R^{2}}\left(1+p^{2R^{2}} \right)+p^{2+2R^{2}} \left(22+p^{2}+3p^{2R^{2}}\right), 
\end{align}
and
\begin{align}
\Omega=&p^{2+R^{2}} \left( 16+24p^{R^{2}}-p^{2+R^{2}}+16p^{2R^{2}}+5p^{3R^{2}}\right)+\notag\\&+p^{2}\left(3+p^{2} \right)+p^{2R^{2}} \left(1-p^{2R^{2}} \right).
\end{align}
Thus, it is clear that having the Jensen-Shannon quantum divergence, one can easily evaluate the quantum coherence by taking it square root (eq.(\ref{QC})).\par
\begin{widetext}
	
	\begin{figure}[hbtp]
		{{\begin{minipage}[b]{.25\linewidth}
					\centering
					\includegraphics[scale=0.355]{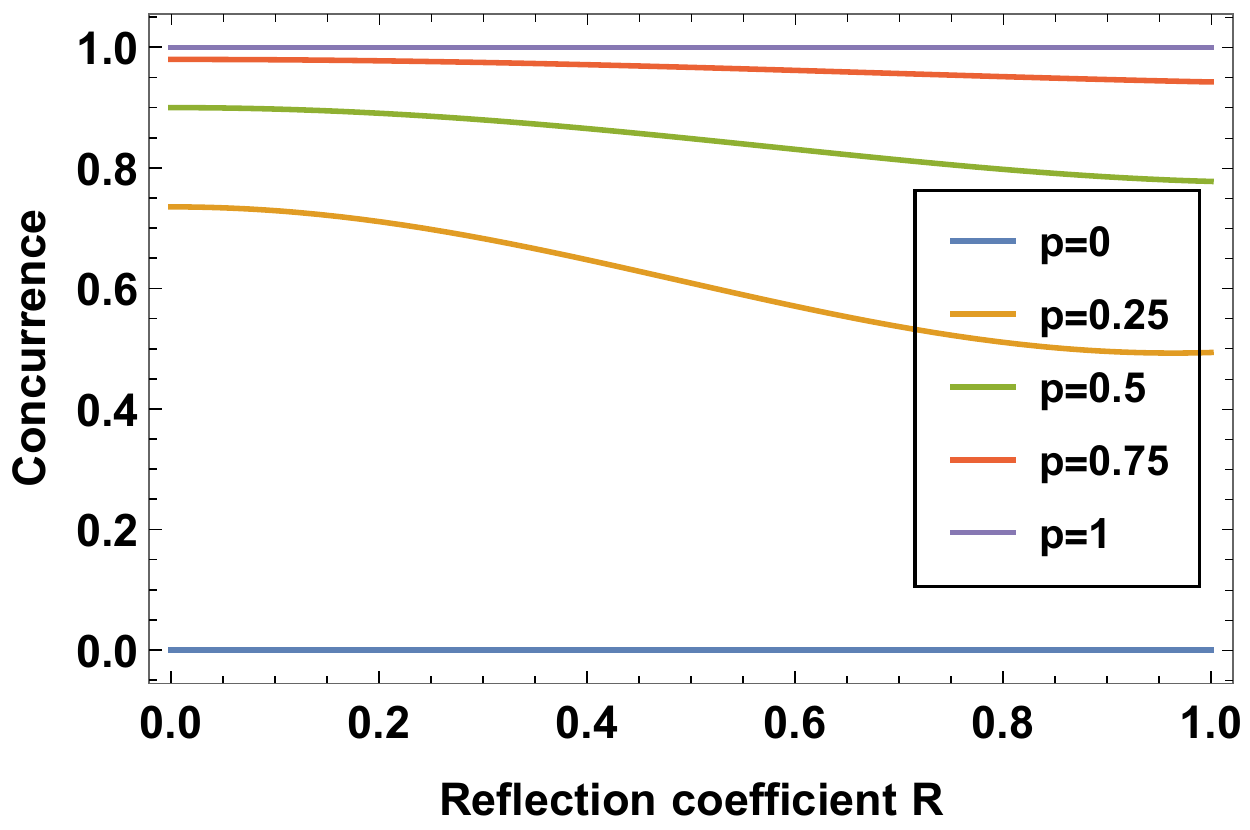} \vfill $\left(a\right)$
				\end{minipage}\hfill
				\begin{minipage}[b]{.25\linewidth}
					\centering
					\includegraphics[scale=0.355]{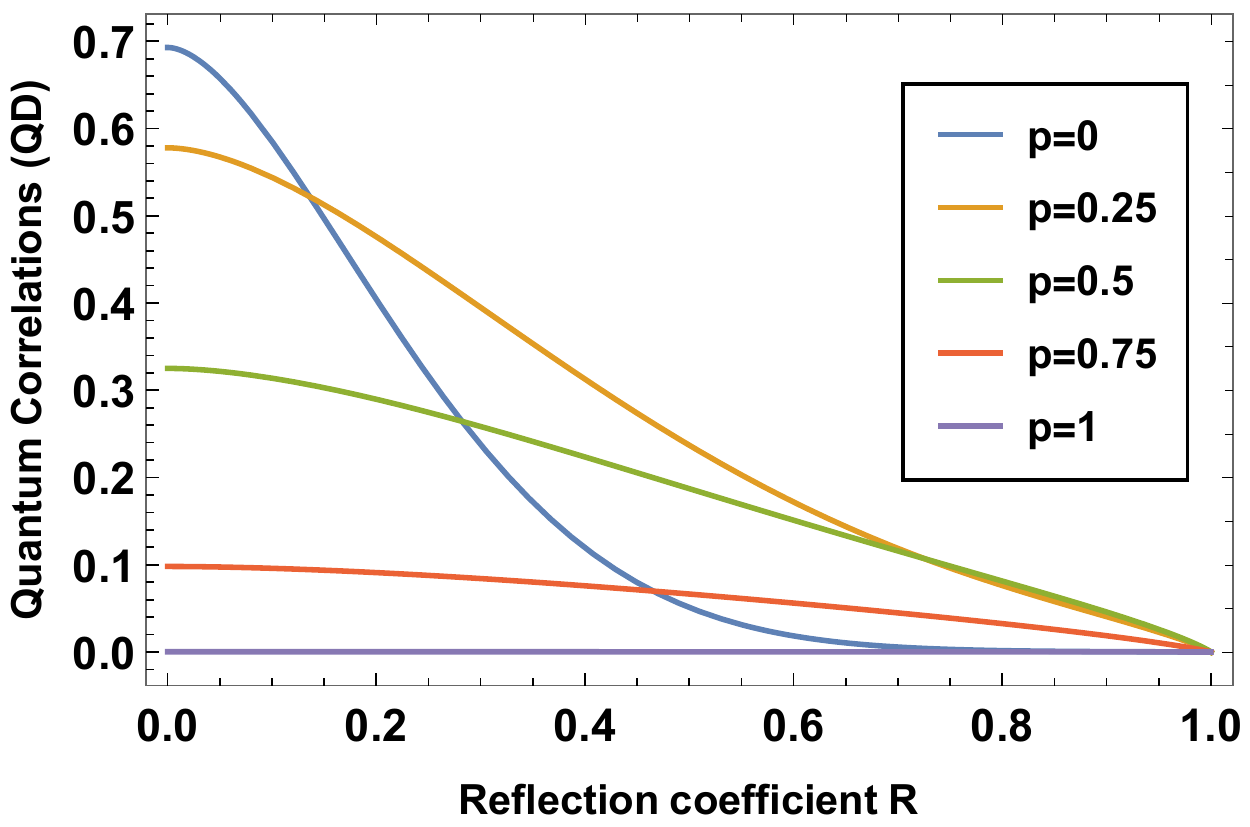} \vfill  $\left(b\right)$
				\end{minipage}\hfill
				\begin{minipage}[b]{.25\linewidth}
					\centering
					\includegraphics[scale=0.355]{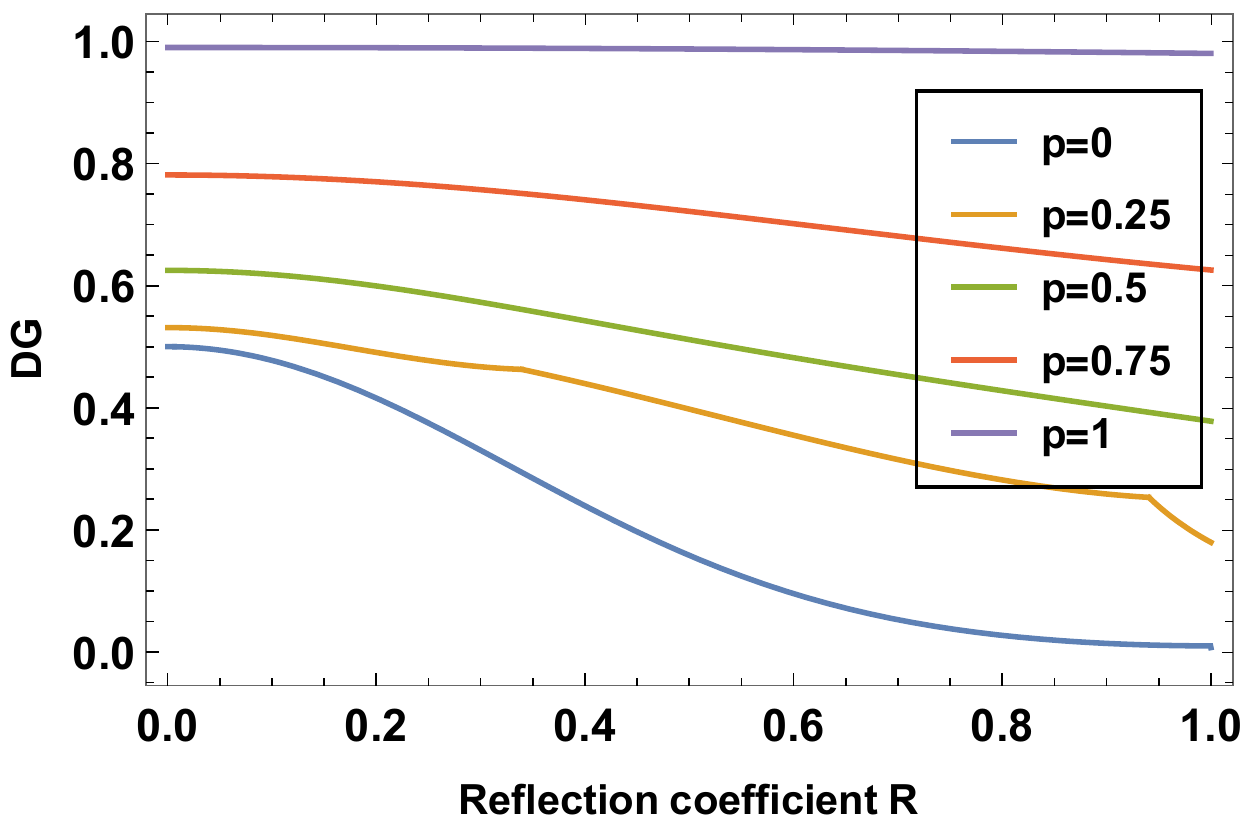} \vfill $\left(c\right)$
				\end{minipage}\hfill
				\begin{minipage}[b]{.25\linewidth}
					\centering
					\includegraphics[scale=0.355]{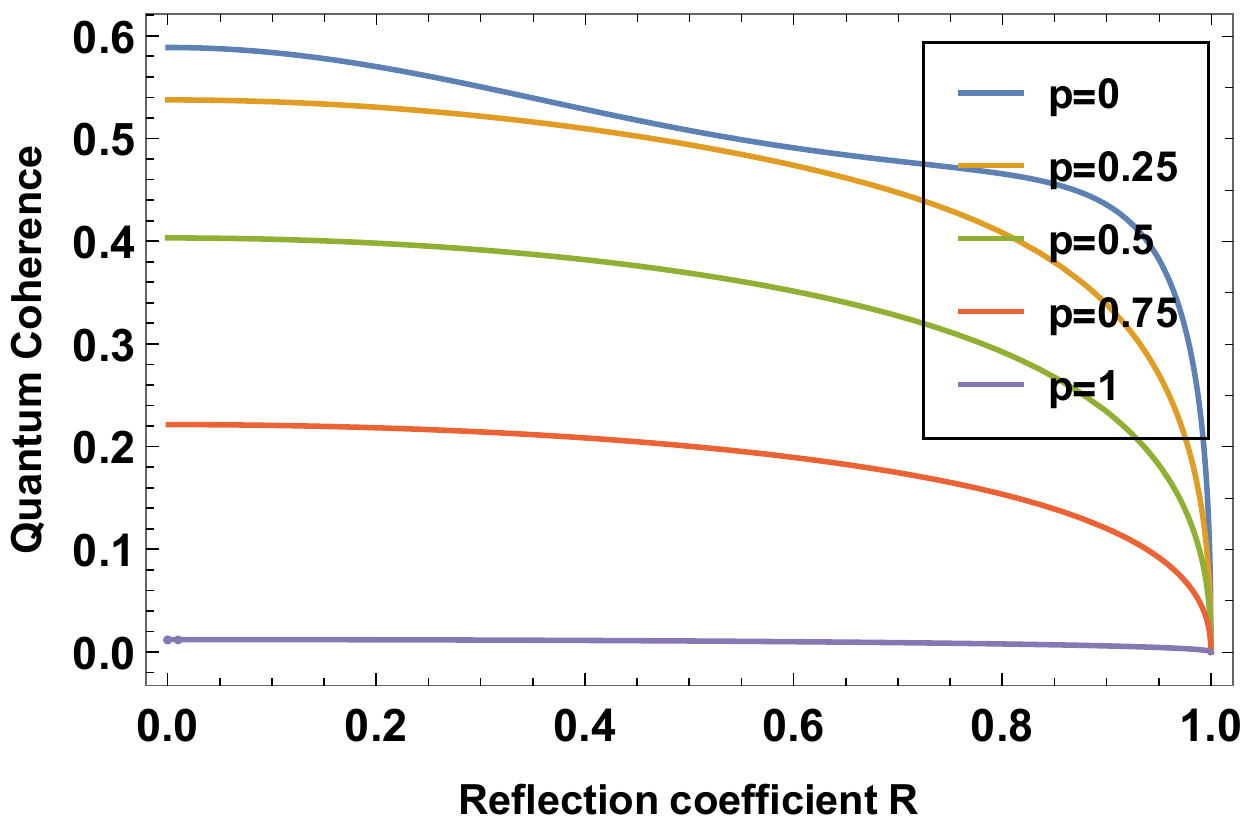} \vfill $\left(d\right)$
		\end{minipage}}}
		\caption{The evolution of concurrence entanglement $1(a)$, entropic quantum discord $1(b)$, trace distance discord $1(c)$ and quantum coherence $1(d)$ versus the reflection coefficient $R$ for different values of the overlapping $p$.}\label{Fig1}
	\end{figure}
\end{widetext}

 
Fig.\ref{Fig1} depicts the behaviors of the the quantum information quantifiers used to characterize the quantum resources in Bell cat states under amplitude damping, versus the reflection coefficient of the beam splitter for particular values of the coherent state overlapping. As we can see form the plot \ref{Fig1}.($a$), the degree of concurrence entanglement of the output state is highly dependent on the overlapping p that specify the coherent state and the reflection coefficient parameter of the beam splitter. Here, the curves show that when the reflection coefficient increases, the Wootter's concurrence decreases and at the same time the depth of the curves decreases as the overlapping values decrease, i.e. the concurrence entanglement increases gradually as $p$ increases. Moreover, one can easily notice that the amount of quantum entanglement achieve almost maximal values when the reflection coefficient of the beam splitter cancels and it does not depend on this coefficient in the limiting cases of overlapping ($p=0,1$), i.e the concurrence remains frozen with respect to the reflection coefficient in these overlap values.\par

The plots in Fig.\ref{Fig1}($b$) illustrates the behavior of the quantum discord based on the von Neumann entropy under the same parameters as in Fig.\ref{Fig1}($a$). We notice that the quantum discord tends slowly towards zero with the increase of the parameter $R$. Actually, it decreases quickly from the maximum reached at $R=0$ to a value beyond which its decrease slows down towards zero for significantly high values of the reflection coefficient, which means that the non-classical correlation tends to disappear in the total reflection of the incident state on an input port of the beam splitter. Contrary to what we have witnessed with entanglement, for the fixed beam splitter reflection parameter, the entropic quantum discord increase as we decreases the coherent state overlapping and its maximum is obtained in the limiting case $p=0$. Interestingly, the freezing behavior with regard to the reflection coefficient is not observable for the entropic quantum discord. This freezing phenomenon exhibited by the Wootters concurrence reflects that the quantum entanglement in a given Bell cat state is not affected by the reflection coefficient parameter of the beam splitter being employed.\par

Let us now analyze the behavior of geometric quantum discord quantified via the Schatten $1$-norm. As depicted in the Fig.\ref{Fig1}($c$), we notice that the plots of the geometric quantum discord have similar behavior as the concurrence, i.e. the non-classical correlation decreases during the action of a beam splitter until it reaches a minimum value in the total reflection $R=1$, which implies that the decoherence effects make the system less correlated. Besides, geometric quantum discord reveals more quantum correlations than concurrence entanglement, and this confirms that the geometric measure of quantum discord using the Schatten norm can exhibit more robustness than concurrence in Bell cat states under amplitude damping. This feature is consistent with the results reported in \cite{Werlang2009} where Werlang and his co-workers showed that quantum discord is more robust than entanglement against decoherence in Markovian environments.\par

In agreement with the results obtained in Fig.\ref{Fig1}($b$), we observe in Fig.\ref{Fig1}($c$) a decrease in the quantum coherence measured by quantum Jensen-Shannon divergence as the reflection parameter goes higher. Here we also remark that quantum coherence decreases with increasing of reflection coefficient as the overlap values decreases. More importantly, the amount of quantum coherence is greater and goes beyond entropic quantum discord. Typically, this can be interpreted by the fact that the total quantum coherence in multipartite systems has contributions from local coherence on subsystems and collective coherence between them, and this result is completely in tune with the physical explanation given in \cite{Yao2015}. From another side, these different measures of quantum criteria behave differently in their evolution.
Now, we want to examine and explain our results in quantum metrology by involving and deciding the sensitivity of the generated Bell cat states as probe states for interferometric phase estimation. To accomplish this, we investigate both LQFI and LQU behaviors and compare them with the behaviors of quantum resources discussed above and further show the role of the latter in improving high precision protocols in which information is encoded in Bell cat states.\par
To illustrate the results of the previous section regarding the analytical expression of the local quantum Fisher information and the local quantum uncertainty based on weak measurements, one computes first both the eigenvalues and the eigenvectors of the density matrix represented by Eq.(\ref{Xform11}). In our case, the non-vanishing eigenvalues take the form
\begin{align}
	\eta_{1,2}=\frac{1\mp q_{R\alpha}}{N_{\alpha}^{\rm Bell}}\left[1\mp p^{2
		-R^{2}} \right],
\end{align}
and consequently, the corresponding eigenstates are given by
\begin{align}
	\left|\varphi_{1}\right\rangle=\frac{1}{\sqrt{a_{\alpha}^{2}a_{\alpha T}^{2}+b_{\alpha}^{2}b_{\alpha T}^{2}}}&\left[a_{\alpha}a_{\alpha T}\left|u_{\alpha},u_{\alpha T} \right\rangle \right. \notag\\&\left. +b_{\alpha}b_{\alpha T}\left|v_{\alpha},v_{\alpha T} \right\rangle\right],
\end{align}
and 
\begin{align}
	\left|\varphi_{2}\right\rangle=\frac{1}{\sqrt{a_{\alpha}^{2}b_{\alpha T}^{2}+b_{\alpha}^{2}a_{\alpha T}^{2}}}&\left[ a_{\alpha}b_{\alpha T}\left|u_{\alpha},v_{\alpha T} \right\rangle \right.\notag\\& \left. +b_{\alpha}a_{\alpha T}\left|v_{\alpha},u_{\alpha T} \right\rangle\right]. 
\end{align}
Using the definition (\ref{QF}), the $3\times3$ symmetric matrix ${\cal M}$ is obtained in the following form
\begin{eqnarray}
	{\cal M}=\left(
	\begin{array}{cccc}
		{\cal M}_{11} & {\cal M}_{12}&0 \\
		{\cal M}_{21} & {\cal M}_{22} &0\\
		0&0&0
	\end{array}
	\right), \label{MatrixM}
\end{eqnarray}
with the elements ${\cal M}_{ij}$ ($i,j=1,2$) are calculated analytically from equation (\ref{Mij}) and they are given by
\begin{equation}
{\cal M}_{11}=\frac{\left( 1-p^{2}\right)\left(1-p^{2R^{2}} \right)}{2\left(1+p^{2} \right)^{2}},
\end{equation}
\begin{equation}
{\cal M}_{22}=\frac{\left( 1-p^{2}\right)\left(1-p^{2R^{2}} \right)p^{2\left(1-R^{2} \right)}}{2\left(1+p^{2} \right)^{2}},
\end{equation}
and
\begin{equation}
{\cal M}_{12}=-{\cal M}_{21}=\frac{i\left( 1-p^{2}\right)\left(1-p^{2R^{2}} \right)p^{1-R^{2}}}{2\left(1+p^{2} \right)^{2}}.
\end{equation}
The expression of local quantum Fisher information (\ref{powerDEF}) is further simplified as
\begin{equation}
\mathcal{Q}_{\cal F}\left(\rho_{AB}\right)=1-\max\left\lbrace Q_{+},Q_{-}\right\rbrace, 
\end{equation}
with the eigenvalues $Q_{\pm}$ of the matrix ${\cal M}$ (\ref{MatrixM}) are given as
\begin{equation}
Q_{\pm}=\frac{1}{2}\left({\cal M}_{11}+{\cal M}_{22}\pm\sqrt{\Delta} \right), 
\end{equation}
with
\begin{equation}
\Delta={\cal M}_{11}^{2}+{\cal M}_{22}^{2}+4{\cal M}_{12}{\cal M}_{21}-2{\cal M}_{11}{\cal M}_{22}.
\end{equation}
On the other hand, using density-matrix elements of equation (\ref{Xform11}) and simplifying equation (\ref{w-elements}), we obtained the eigenvalues of the $3\times3$ symmetric matrix $W$. It is easy to check that
\begin{align}
w_{11}=\frac{\sqrt{\xi}}{2\left(p^{2}+1\right)}&\left[1+\frac{1}{\xi}\left[\left(1-p^{2}\right)\left(1-p^{2R^{2}} \right)\left(1-p^{2\left(1-R^{2} \right)}\right) \right.\right. \notag\\&\left.\left. +  \left( p^{1-R^{2}}+p^{1+R^{2}}\right)-4p^{2} \right]  \right],\label{w11}
\end{align}
\begin{align}
	w_{22}=\frac{\sqrt{\xi}}{2\left(p^{2}+1\right)}&\left[1+\frac{1}{\xi}\left[\left( p^{1-R^{2}}+p^{1+R^{2}}\right)-4p^{2} \right.\right. \notag\\&\left.\left. -\left(1-p^{2}\right)\left(1-p^{2R^{2}} \right)\left(1-p^{2\left(1-R^{2} \right)}\right) \right]  \right],\label{w22}
\end{align}
where
\begin{align}
\xi=\left(1-p^{2R^{2}} \right)\left(1-p^{2\left(2-R^{2} \right) } \right),  
\end{align}
and
\begin{align}
w_{33}=\frac{1}{2}+\sum_{\pm}&\left[ \frac{\left(p^{1-R^{2}}+p^{1+R^{2}}\pm2p \right)^{2}}{4\left(1\pm p^{R^{2}} \right)\left(1\pm p^{2-R^{2}} \right)\left( p^{2}+1\right)}\right. \notag\\&\left. -\frac{\left(1-p^{2} \right)\left(1\pm p^{R^{2}} \right)^{2} \left(1-p^{2\left(1-R^{2} \right) } \right)}{4\left(1\pm p^{R^{2}} \right)\left(1\pm p^{2-R^{2}} \right)\left( p^{2}+1\right)}\right].\label{w33} 
\end{align}
Thus, an analytical expression of the LQU can be determined by substituting equations (\ref{w11}), (\ref{w22}) and (\ref{w33}) into equation (\ref{LQU1}). To proceed, from above equations it is clear that $w_{11}\geq w_{22}$. Therefore, the compact formula of the weak measurement-induced local quantum uncertainty (\ref{UW}) turns out to be
\begin{equation}
\mathcal{U}_{W}\left(\rho_{AB}\right)=\frac{\cosh\left(\chi\right)-1}{4\cosh\left(\chi\right)}\left( 1 - \max\{w_{11},w_{33}\}\right). \label{UWW}
\end{equation}
\begin{figure}[hbtp]
	{{\begin{minipage}[b]{.45\linewidth}
				\centering
				\includegraphics[scale=0.35]{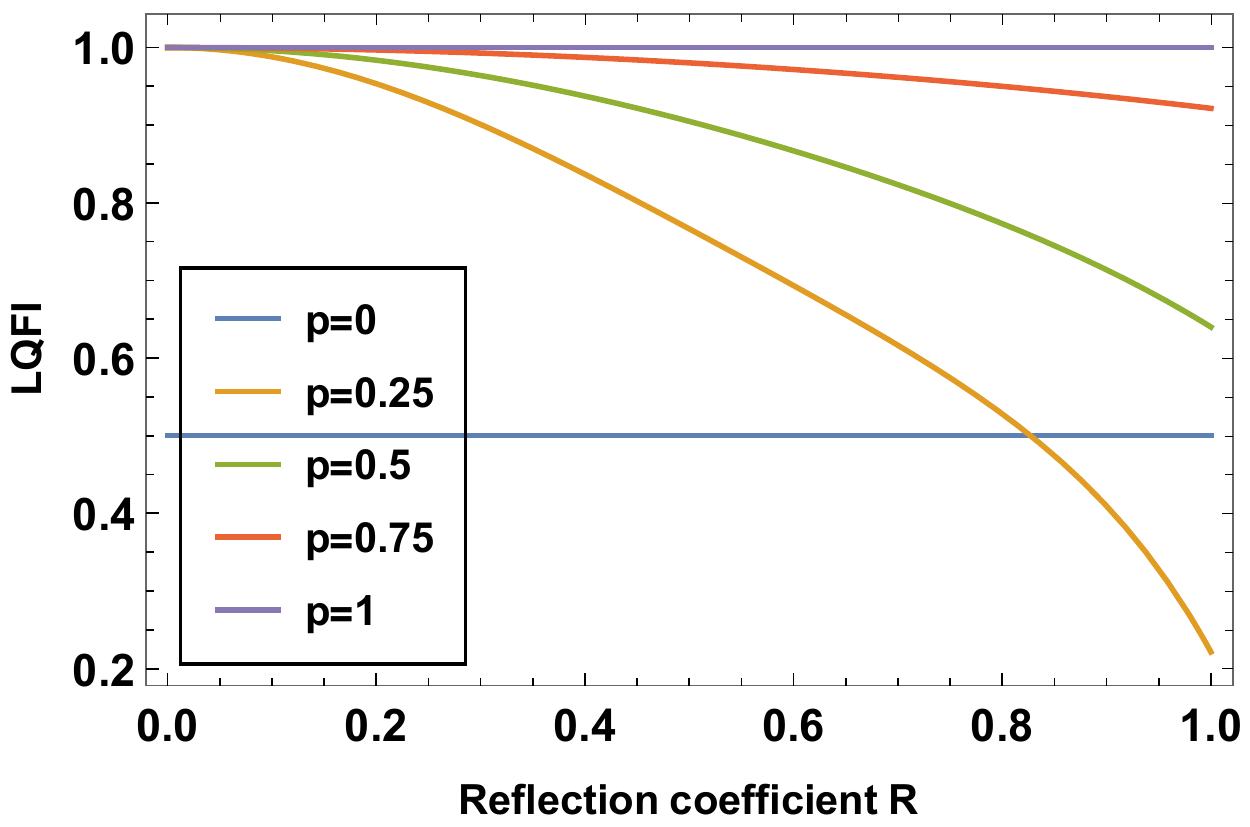} \vfill $\left(a\right)$
			\end{minipage}\hfill
			\begin{minipage}[b]{.48\linewidth}
				\centering
				\includegraphics[scale=0.35]{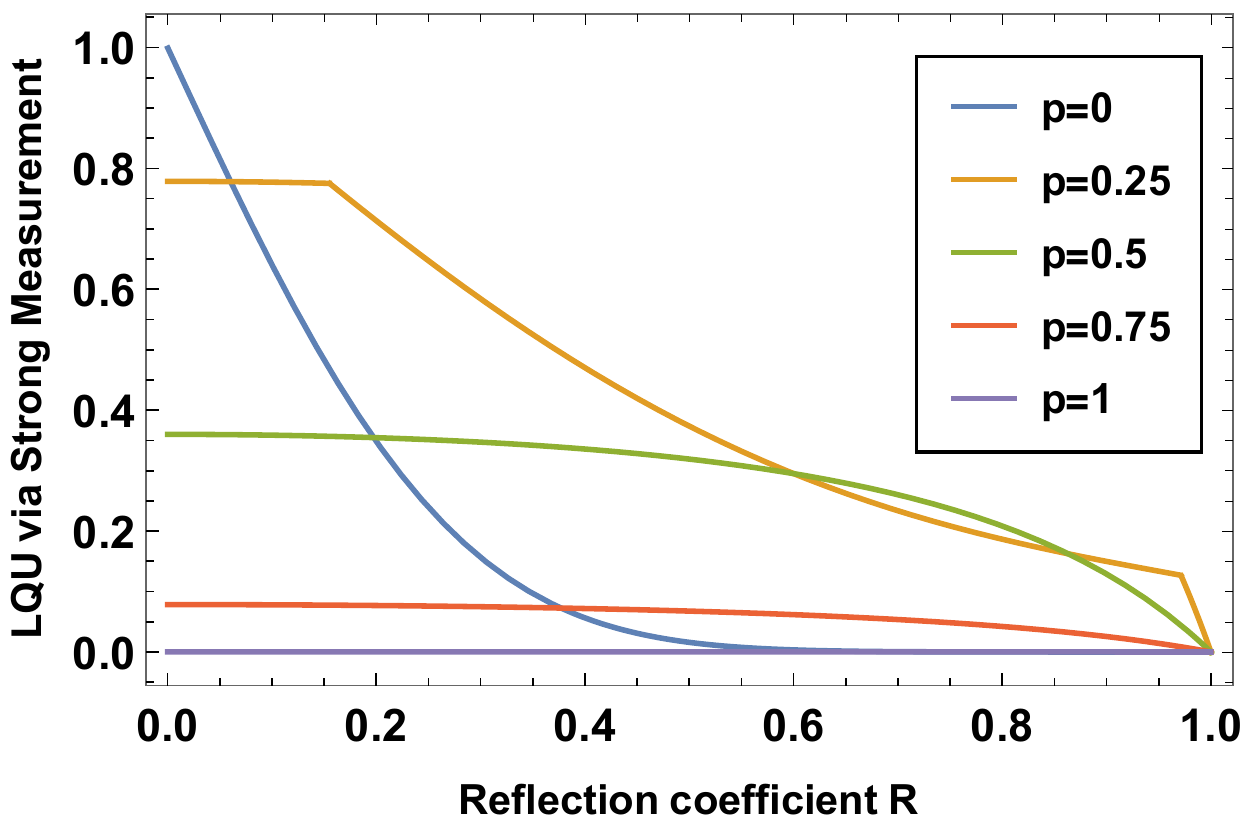} \vfill  $\left(b\right)$
	\end{minipage}}}
	\caption{($a$) The evolution of the local quantum Fisher information of Bell cat states under amplitude damping channel with respect to the beam splitter reflection coefficient for given coherent state overlapping $p$. ($b$) Similar to ($a$) but for the local quantum uncertainty with the projective measurements.}\label{Fig2}
\end{figure}

Using these analytical results, we depict in Fig.\ref{Fig2} the metrological measures of the non-classical correlations at different reflection coefficient of a unitary beam splitter for various values of coherent state overlap. For higher values of the overlapping $p$, the Fig.\ref{Fig2}($a$) shows that the local quantum Fisher information reduced by an increase of the reflection coefficient, and it vanishes at higher reflection parameter. LQFI is equal to one at $R=0$ and only starts to decrease after a reflection threshold of the incident state in the beam splitter. Further, LQFI is more robust against the amplitude damping effect in the limiting cases of overlapping $p=0,1$. As well, we note that increasing the overlap degree between the two coherent states leads to an increase in the LQFI. From the results reported in Fig.\ref{Fig2}($b$), we can clearly see that LQU and LQFI have the same exponential decay behavior, except that here LQU decreases with increasing degree of coherent state overlapping.\par

By comparing Fig.\ref{Fig2}($a$) and Fig.\ref{Fig2}($b$), we conclude that the amount of non-classical correlations captured by LQFI in Bell cat states is larger than those of LQU and that its amplitude is still much larger. This can be interpreted using the relationship between the Wigner-Yanase skew and the quantum-Fisher information quantifiers reported in \cite{Slaoui22019}, $\mathcal{U}\leqslant\mathcal{Q}_{\cal F} \leqslant2\mathcal{U}$, which reflects the fact that LQU is majorized by LQFI in any metrological task of phase estimation and that LQU values are always smaller than LQFI values in qubit-qudit systems. It is worth emphasizing that LQFI and LQU are closely related to the error boundary of the parameter estimation, namely, $var\left(\theta\right)_{\min}\leqslant\mathcal{Q}_{\cal F}^{-1}$ and $var\left(\theta\right)_{\min}\leqslant\mathcal{U}^{-1}$. Hence, more precision in any interferometric phase protocol is obtained for the largest value of the both LQU and LQFI. Thus, these non-classical correlation measures based on the notion of quantum uncertainty are useful in parameter estimation compared to other measures previously discussed. \par

Generally speaking, it is well known that non-classical correlations are being used to enhance the precision of phase estimation, and thus, the information gain becoming robust to noise is critical to promising these non-classical properties in quantum metrology \cite{Petz2011}. Indeed, even better accuracy can often be achieved by using the correlated states as the output probe state in the interferometric phase estimation scheme. In our generated Bell cat states (\ref{Xform11}), the maximum value of the LQFI is achieved when $p\longrightarrow1$. This indicates that the corresponding states hold the highest quantum correlations and the best interferometric phase estimate is obtained. Besides, the maximum values of both LQU and LQFI are reached for the total transmission of the incident state on an input port of the beam splitter, that is, in the limiting case wherein $R=0$. This limit provides the best estimation efficiency and to obtain the optimal estimation efficiency, it is important to employ the Bell cat states with $R=0$ and $p=1$.

\begin{widetext}
	
\begin{figure}[hbtp]
	{{\begin{minipage}[b]{.25\linewidth}
				\hfill $\left(a\right)$ \centering 
				\includegraphics[scale=0.35]{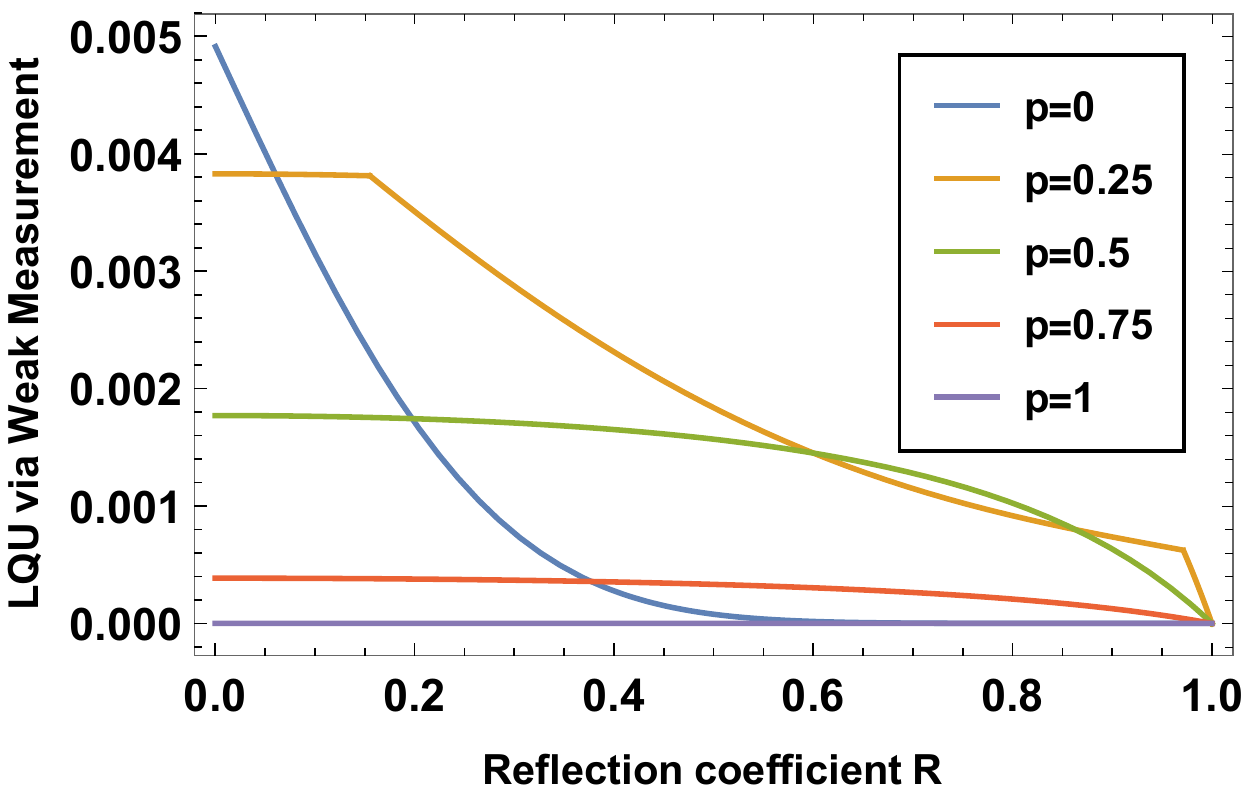} \vfill Measurement Strength $\chi=0.2$
			\end{minipage}\hfill
			\begin{minipage}[b]{.25\linewidth}
				\centering \hfill $\left(b\right)$
				\includegraphics[scale=0.35]{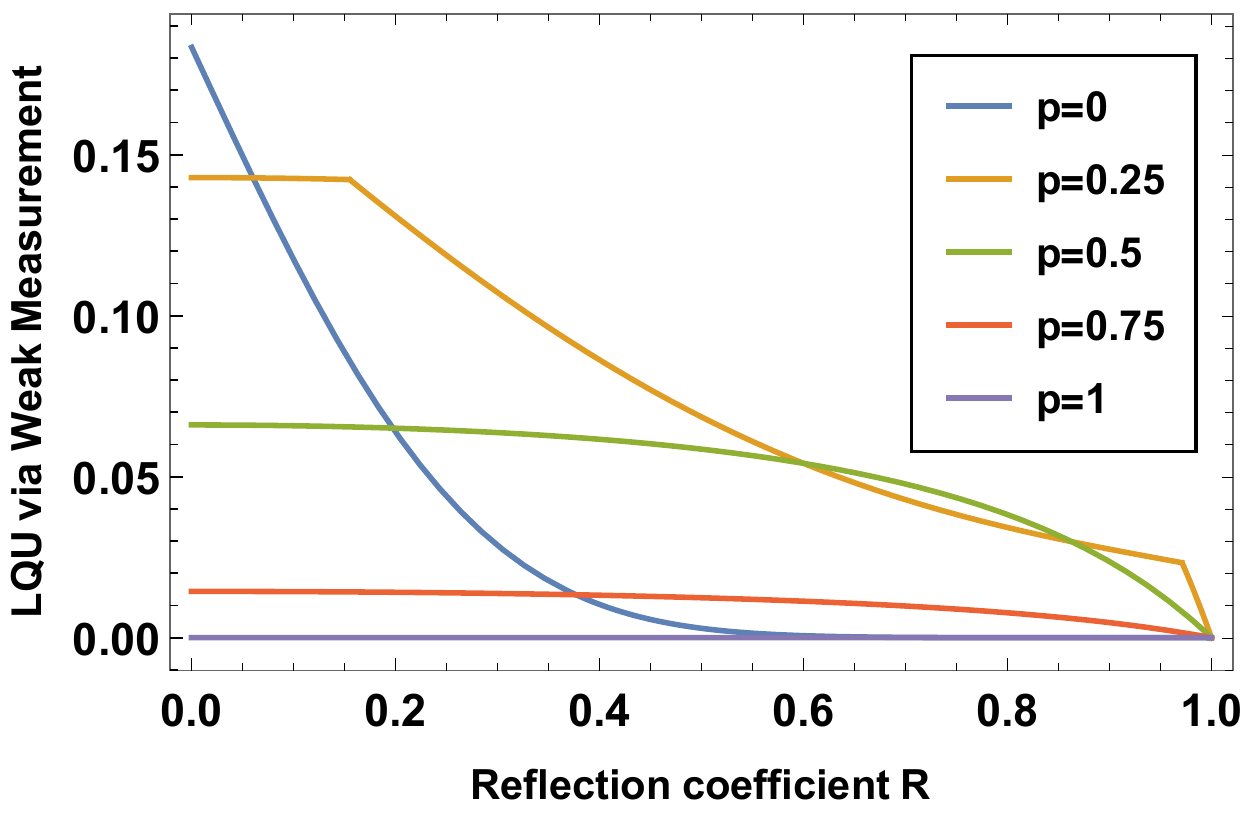} \vfill Measurement Strength $\chi=2$
			\end{minipage}\hfill
			\begin{minipage}[b]{.25\linewidth}
				\centering \hfill $\left(c\right)$
				\includegraphics[scale=0.35]{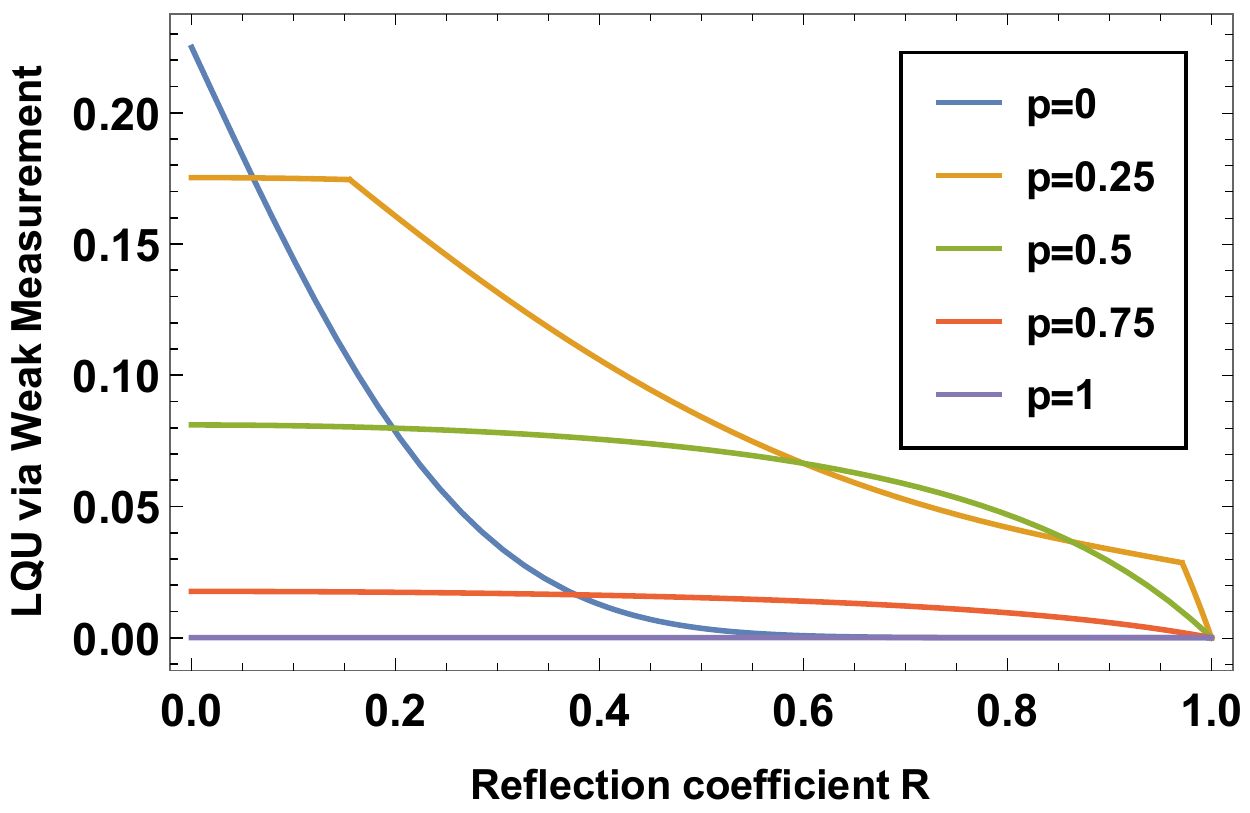} \vfill Measurement Strength $\chi=3$ 
	\end{minipage}\hfill
\begin{minipage}[b]{.25\linewidth}
\centering \hfill $\left(d\right)$ 
\includegraphics[scale=0.35]{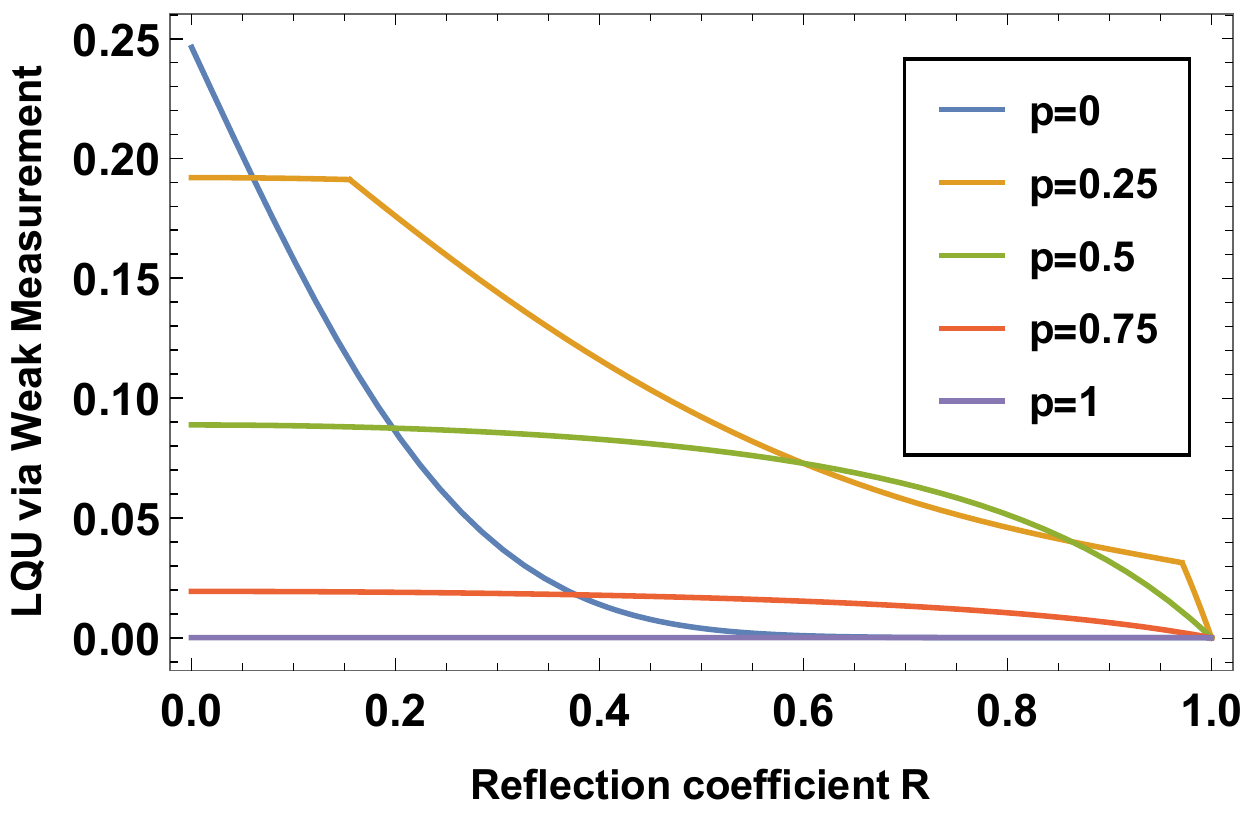} \vfill Measurement Strength $\chi=5$
\end{minipage}}}
	\caption{Weak measurement-induced local quantum uncertainty as a function of the reflection coefficient $R$ for different values of the overlapping $p$, which can be numerically verified using Eq.(\ref{UWW}). Fig.($a$) is obtained for the measurement strength $\chi=0.2$, Fig.($b$) is for $\chi=2$, Fig.($c$) is for $\chi=3$ and Fig.($d$) is for $\chi=5$.}\label{Fig3}
\end{figure}
\end{widetext}

In order to see the effect of the measurement strength on quantum correlation quantifier LQU, we plot in Fig.\ref{Fig3} the weak measurement-induced LQU versus the reflection coefficient for various values of overlapping and measurement strength $\chi$. As can be seen, we remark that LQU by the projective measurements (see Fig.\ref{Fig2}($b$)) is larger than the LQU induced by the weak measurement (i.e., $\mathcal{U}\geq\mathcal{U}_{W}$). For comparison, weak LQU approaches to zero for smaller values of  the measurement strength x, and can be approaching the normal LQU in the limit of strong projective measurement $\chi\longrightarrow\infty$. Moreover, weak measurement-induced LQU and normal LQU have the same decay behavior and the only difference is their respective amplitudes. Our results indicate that weak measurements do not always reveal more quantumness than strong measurements as reported in \cite{Singh2014}. Indeed, weak measurements weakly perturb the subsystem of the composite system, so the entropy of the composite system does not change much and thus the uncertainty of measurement is small.
\section{Concluding Remarks and Outlook}\label{V}
Since they identify the states needed for certain tasks, quantum resource theories are crucial for quantum information processing. For instance, it has been shown that entangled states are necessary to achieve a quantum advantage for a variety of protocols. Further, QE, QD, and other quantum resources must be preserved for a longer period during such protocols. This might be achieved by researching various external transmitting mediums and characterizing the related controlled parameters. In the present work, we explored the quantumness of the Bell coherent-state superpositions produced by a beam splitter device, where the vacuum state is incident on one input port and the Glauber coherent state is incident on the other. Different quantifiers are used to measure the quantumness in the output state and to obtain the analytical expression of concurrence entanglement, entropic quantum discord, trace norm quantum discord, quantum Jensen-Shannon divergence, local quantum uncertainty and local quantum Fisher information for generated states under the amplitude damping channel. A beam splitter's action qualitatively reflects decoherence effects, and the evolution of these quantumness are heavily influenced by the coherent state overlapping and the beam splitter's reflection coefficient parameter. This implies that quantum resources in physical systems based on an optical beam splitter device could be controlled by adjusting these parameters.\par

On the other hand, by using continuously infinitesimal weak measurements in conjunction with an ideal projective measurement, the quantumness exhibited in the quantum systems can be recovered. Here, we have provided an explicit analytical form of local quantum uncertainty via weak measurements, which applies to any qubit-qudit quantum system. However, the global amount recovered via weak measurements is lower than the one extracted by projective measurements. In fact, we have found that the weak measurement-induced LQU is lower than normal LQU captured by projective measurement for smaller values of  measurement strength and they approach each other for the larger values. This explains why projective measurements cannot be achieved by continuous infinitesimal weak measurements. Unlike projective measurements, which completely extract the quantumness from a quantum system and convert it into a classical system, weak measurements extract only a very small part of quantum correlation.\par

We further examine the role of pairwise quantum correlations when the generated Bell coherent state superpositions are used as a probe state in estimation protocols for an unknown phase shift. By controlling the reflection coefficient of a beam splitter, it is clear that an increase in coherent state overlap induces a decrease in the metrological measures of the non-classical correlations (LQU and LQFI), and that they are more robust against the amplitude damping effect in the limiting case of overlapping ($p=1$), wherein interestingly the best accuracy in the interferometric phase estimation is given at total transmission ($R=0$). Furthermore, we show that the sensitivity of the phase estimation depends on the strength with which the probe state is perturbed, where increasing the measurement strength leads to an optimal estimation. This investigation on the significance of non-classical correlations and weak measurement in quantum metrology may inspire further experimental work aimed at developing highly accurate protocols that encode information in Bell coherent-states superpositions.

\end{document}